\documentclass[%
reprint,
superscriptaddress,
 amsmath,amssymb,
 aps,
floatfix,
]{revtex4-1}

\usepackage{graphicx}% Include figure files
\usepackage{dcolumn}% Align table columns on decimal point
\usepackage{bm}% bold math
\usepackage{color}
\usepackage{slashed}
\usepackage{comment}
\usepackage{tikz-cd}
\usepackage{mathtools}

\usepackage[linktocpage,colorlinks=true,linkcolor=blue,citecolor=blue,urlcolor=blue,breaklinks=true]{hyperref}   %Perfect hyperlink

\def\ave#1{\langle \psi |#1 | \psi \rangle}
\def\Ave#1{\langle #1 \rangle}
\def\sp#1{\mathrm{Span}(#1)}
\def\Re{\mathrm{Re}}
\def\Im{\mathrm{Im}}
\def\fpr{f_{\mathrm{pr}}}
\def\faf{f_{\mathrm{aff}}}
\def\J{\mathcal{J}}
\def\Ls{\mathcal{L}_s(\mathcal{H})}

\begin{document}

%\preprint{APS/123-QED}

\title{Quantum Geometry of Expectation Values}% Force line breaks with \\

\author{Chaoming Song}
\email{c.song@miami.edu}
\affiliation{%
Department of Physics, University of Miami, Coral Gables, Florida 33146, USA}%



%\date{}% It is always \today, today,

\begin{abstract}
We propose a novel framework for the quantum geometry of expectation values over arbitrary sets of operators and establish a link between this geometry and the eigenstates of Hamiltonian families generated by these operators.  We show that the boundary of expectation value space corresponds to the ground state, which presents a natural bound that generalizes Heisenberg's uncertainty principle. To demonstrate the versatility of our framework, we present several practical applications, including providing a stronger nonlinear quantum bound that violates the Bell inequality and an explicit construction of the density functional. Our approach provides an alternative time-independent quantum formulation that transforms the linear problem in a high-dimensional Hilbert space into a nonlinear algebro-geometric problem in a low dimension, enabling us to gain new insights into quantum systems.
\end{abstract}

\maketitle

%\tableofcontents

\section{Introduction}\label{sec:intro}

Boundedness is a ubiquitous phenomenon in the quantum realm. One of the most prominent examples is Heisenberg's uncertainty principle \cite{Heisenberg1927,kennard1927quantenmechanik,weyl1950theory}, which sets a nonlinear bound on the expectation values of quadratic position and momentum operators. Another well-known example is the Tsirelson bound \cite{cirel1980quantum} for quantum nonlocality, which limits the ability of quantum mechanics to violate Bell inequalities \cite{brunner2014bell}. Density functional theory (DFT) \cite{parr1980density} is another notable example, where two theorems introduced by Hohenberg and Kohn (HK) \cite{hohenberg1964inhomogeneous} state that the system energy is bounded by a unique functional of the particle density and is uniquely saturated by the ground state. The HK functional depends only on the interactions, making it universal and applicable to any electron system. However, unlike Heisenberg's uncertainty principle, the exact form of the HK functional remains unknown \cite{kohn1965self,parr1980density,geerlings2003conceptual,becke2014perspective,koch2015chemist}, despite limited efforts of numerical searches \cite{levy1979universal,levy1982electron,lieb1983density}.

Despite many isolated examples of quantum boundedness regarding expectation values, there is currently no general framework that unifies these cases into a coherent whole. In particular, the textbook generalization of Heisenberg's uncertainty relation is limited to expectation values of two operators and their commutator, and is therefore incapable of applying to many other problems. For instance, in the Bell setup, one needs to examine the expectation values of four nonlocal operators. Furthermore, the particle density in DFT is typically an infinite set, making the problem even more challenging.

In this research article, we present a comprehensive framework for the quantum geometry of expectation values over an arbitrary set of operators. Our framework draws inspiration from a fundamental observation: the singular set of the mapping from a Hilbert space to the space of expectation values of a given set of operators is associated with the eigenstates of the Hamiltonian family generated by these operators, in accordance with the variational principle. In particular, as a subset of this singular set, the boundary of the expectation space corresponds to the ground states, which provides a non-trivial bound of quantum expectations. 

We demonstrate Heisenberg's uncertainty relation as a special case of our theory, thereby generalizing the certainty principle. Furthermore, the  geometry of the expectation values determines time-independent quantum theories completely, without invoking the quantum state. This potentially leads to the development of an alternative quantum formulation that transforms the linear problem in a high-dimensional Hilbert space into a nonlinear algebro-geometric problem in a lower dimension.

Section~\ref{framework} outlines the general theoretical framework of our approach. Specifically, in Subsec.~\ref{sec:moduli}, we introduce the expectation moduli space as the space formed by the expectation values and the singular moduli as its singular subset, and establish their relations with the eigenstates and the ground states. In Subsec.~\ref{sec:polynomial}, we construct the projective dual of the singular moduli and connect it to the characteristic polynomial. In Subsec.~\ref{sec:HUP}, we show that Heisenberg's uncertainty relation can be derived directly from our framework, illustrating its generality and usefulness. Subsec.~\ref{sec:degenerate} discusses the case of degeneration and the moduli space of integrable models. We then discuss the relationship between all moduli spaces in Subsec.~\ref{sec:cat}, arguing that our framework offers an alternative quantum formulation. In Subsec.~\ref{sec:classical}, we discuss the classical counterpart of expectation moduli, and demonstrate the semiclassical construction and its connection to Gutzwiller's trace formula. Finally, in Subsec.~\ref{sec:FT}, we discuss a finite-temperature analogy of our approach and the potential generalization to quantum field theory. 
 
Section~\ref{application} presents a range of potential applications of our framework. Specifically, in Subsec.~\ref{sec:bell}, we apply our theory to the study of quantum nonlocality, which leads to a new quantum bound that is stronger than the Tsirelson bound. In Subsec.~\ref{sec:DFT}, we provide a detailed examination of how our theory applies to DFT. Finally, in Subsec.~\ref{sec:NRP}, we apply our theory to the $N$-representability conditions for the reduced density matrices. 

\section{General Framework}\label{framework}

\subsection{Expectation value moduli}\label{sec:moduli}

\begin{figure}
 \includegraphics[width=1\linewidth]{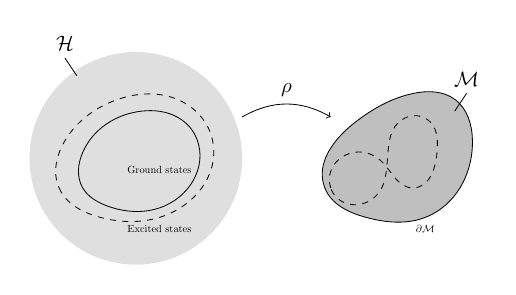}
 \caption{ Mapping $\rho$ from the Hilbert space $\mathcal{H}$ (grey domain) to the moduli space  $\mathcal{M}$ (dark grey domain). The solid and dashed curves in $\mathcal{H}$ correspond to the ground and excited states, where the former maps to the boundary $\partial \mathcal{M}$.
 }. 
 \label{fig:map}
\end{figure}

Consider a Hilbert space $\mathcal{H}$ and the corresponding real vector space of self-adjoint operators, $\Ls$. To simplify our analysis, we assume that $\mathcal{H}$ has a finite dimension $N$. Our focus is on a particular operator space $O \subseteq \Ls$ of dimension $M+1$, which includes the identity operator $I$ as one of its elements. We choose a set of linearly independent basis operators $\mathbf{H} \coloneqq \{ H_i\}$, and any operator in this space can be expressed as a linear combination of $\mathbf{H}$. We are interested in the geometry of the expectation values of $ \mathbf{H}$, which motivates the introduction of a mapping $\rho$ from $\mathcal{H}$ to a $M$-dimensional real projective space $\mathbb{RP}^M$. Specifically, we define 
 \begin{equation}
    \rho(\psi, \psi^\dag)  \coloneqq \left( \ave{H_0} , \ldots, \ave{H_{M}} \right), 
\label{eq:rho}
 \end{equation}
using a set of coordinates $(\rho_0, \ldots, \rho_M)$, where $ \rho_i = \ave{H_i} $ denotes the expectation value of an operator. The moduli space of expectation values is determined by the image of the mapping $\rho$, denoted as $\mathcal{M}(O) \coloneqq \textrm{im}(\rho)$, which is generally a semialgebraic set. In most of the discussion below, we will fix the operator space $O$ and omit it from the notation of $\mathcal{M}$. Figure~\ref{fig:map} illustrates the mapping $\rho$ from $\mathcal{H}$ to $\mathcal{M}$. It is noteworthy that the moduli space $\mathcal{M}$ is convex. However, we will not delve into the details of this topic in this paper and refer the interested reader to our previous work \cite{song2023zero}.

The mapping $\rho$ is not injective, meaning that a point in the moduli space $\mathcal{M}$ may correspond to multiple $\psi \in \mathcal{H}$. To gain a better understanding of this, we consider the invariant class of infinitesimal changes $\psi \rightarrow \psi + \delta \psi$ that keep $\rho$ unchanged. This condition requires that $\delta \rho = 0$, or equivalently,
\begin{equation} \label{eq:jacobi}
\begingroup
\renewcommand*{\arraystretch}{1.3}
 \mathcal{J}(\psi, \psi^\dag) \begin{bmatrix} \delta \psi \\ \overline{\delta \psi} \end{bmatrix}= 0,
\endgroup
\end{equation}
where $\mathcal{J}(\psi, \psi^\dag) $ is the $M+1 \times 2N$ Jacobian matrix of the mapping~({\ref{eq:rho}}), given by
\begin{equation}\label{eq:J}
 \mathcal{J}(\psi, \psi^\dag)  \coloneqq \left [ \psi^{\dag} H_i  ,  \psi^t  H_i ^t \right],
\end{equation}
where the square bracket represents the row concatenation for all $0\leq i \leq M$. 

Equation~(\ref{eq:jacobi}) shows that the kernel of the Jacobian~(\ref{eq:J}) corresponds to the invariant tangent subspace at the point $(\psi, \overline \psi)$, with dimension $\dim \mathrm{ker} (\mathcal{J}) = 2N -(M+1)$ for generic points. However, singular points, such as the boundary of $\mathcal{M}$, have extra degrees of freedom \cite{hartshorne2013algebraic}, i.e., $\dim \mathrm{ker} (\mathcal{J}) > 2N-(M+1)$. Figure~\ref{fig:boundary} illustrates a toy example of this argument. Geometrically, the set of singularities $\mathcal{M}_s \subseteq \mathcal{M}$ forms a projective hypersurface, whose physical interpretation will become clear later. In particular, the boundary $\partial \mathcal{M}$ is a subset of $\mathcal{M}_s$, i.e., $\partial \mathcal{M} \subseteq \mathcal{M}_s$.

Moreover, the index theorem states that $\dim \mathrm{ker} ({\mathcal{J}}) - \dim \mathrm{coker} ({\mathcal{J}}) = 2N-(M+1)$, indicating that singularities correspond to 
\begin{equation} \label{eq:coker0}
    \dim \mathrm{coker} ({\mathcal{J}})>0.
\end{equation}
Equation~(\ref{eq:coker0}) implies the existence of a non-trivial cokernel $\lambda$ of the Jacobian,
\begin{equation}  \label{eq:coker}
\lambda^t \mathcal{J}(\psi, \psi^\dag) = 0,
\end{equation}
which corresponding to singular set of the mapping~(\ref{eq:rho}). Equation~({\ref{eq:coker}) implies that geometry of singularity set $\mathcal{M}_s$ corresponds to the existence of a cokernel $\lambda$ for the Jocobian $\mathcal{J}$. 

In the case where $M+1 > 2N$, the cokernel condition~(\ref{eq:coker}) is trivially satisfied, which implies that $\mathcal{M}_s = \mathcal{M}$. This is because the operator space has a sufficiently large dimension that the Hilbert space $\mathcal{H}$ can be embedded into $\mathbb{RP}^M$ without a boundary, indicating that $\mathcal{M}$ is an algebraic variety. However, for $M+1 \leq 2N$, Eq.~(\ref{eq:coker0}) becomes non-trivial, and $\mathcal{M}$ becomes a semi-algebraic set with $\mathcal{M}_s$ being its strict subset, i.e., $\mathcal{M}_s \subsetneq \mathcal{M}$. Therefore, unless an explicit statement is provided, we will assume that $M+1 \leq 2N$ from now on.

\begin{figure}
  \includegraphics[width=1\linewidth]{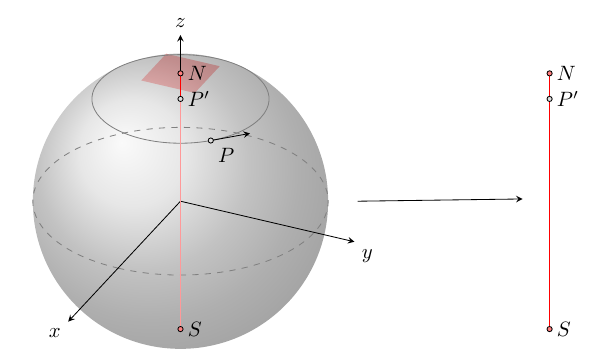}
  \caption{Projection of $S^2$ surface to the line segment $[-1,1]$ on the $z$-axis through $(\theta, \phi)\rightarrow \cos\phi$. The invariant infinitesimal change $0 = \delta z =  -\sin\phi \delta \phi$ requires $\delta \phi = 0$ except for the two poles with $\phi = 0$ or $\pi$. Point $P$, along with points of the same latitude (solid circle), is projected to a single point $P'$. The invariant tangent subspace $\delta \phi = 0$ at point $P$ is one-dimensional, represented by the solid arrow starting from $P$. In contrast, the invariant tangent subspaces of the north and south poles (marked by a red square) are two-dimensional, meaning they map onto the boundaries of the projected space $[-1,1]$.}
  \label{fig:boundary}
\end{figure}

The cokernel condition in Eq.~(\ref{eq:coker}) for the singularity set $\mathcal{M}_s$ has a clear physical interpretation. Consider a family of systems parameterized by a $M$-dimensional real vector $\lambda \in \mathcal{RP}^{M*}$, where each system has a Hamiltonian $H(\lambda) \in O$. We express $H(\lambda)$ as a linear combination of the basis $\mathbf{H}$, such that $H(\lambda) = \sum_{i} \lambda_i H_i$. We define the energy functional $E[\psi]$ as the expectation value of the Hamiltonian, as
\begin{equation}\label{eq:E}
    E[\psi] = \langle   H(\lambda) \rangle  =  \lambda^t   \rho .
\end{equation}
The variational principle suggests that the stationary points of Eq.~(\ref{eq:E}) correspond to the eigenstates $\psi$ of $ H$, satisfying
\begin{equation}\label{eq:var}
0 =   \lambda^t  \frac{\delta \rho}{\delta \psi}   = 
\lambda^t \mathcal{J},
\end{equation}
 where $\mathcal{J} \coloneqq \frac{\delta \rho}{\delta \psi}$ is the Jacobian of the mapping~$\rho$ defined in Eq.~(\ref{eq:J}). is the Jacobian of the mapping $\rho$ defined in Equation~(\ref{eq:J}). Remarkably, this variational equation is precisely equivalent to the cokernel condition~(\ref{eq:coker}). The same conclusion can also be reached directly from the Schr\"odinger equation
\begin{equation}\label{eq:schrodinger}
     (\lambda_0 H_0+\ldots+\lambda_M H_M) \psi = 0,
\end{equation}
where the eigenvalue term has been absorbed by a linear combination of basis. Rather than solving for the eigenstates $\psi$ of a fixed parameter set $\lambda$, we consider the dual problem of solving for $\lambda$ given a fixed $\psi$. This allows us to rewrite the Schr\"odinger equation as $\lambda^t \mathbf{ H} \psi = 0$. Since $\lambda$ is real, we also require its complex conjugate equation, which again recovers the cokernel condition~(\ref{eq:coker}). Note that the Schrödinger equation~(\ref{eq:schrodinger}) provides a sufficient condition for the cokernel condition~(\ref{eq:coker}). However, it does not necessarily guarantee its validity. 

The analysis presented above indicates that the space $\mathcal{M}_e$, which is formed by the expectation values $\{\Ave{\psi_e(\lambda)|H_i|\psi_e(\lambda)}\}$ for all eigenstates of $\psi_e(\lambda)$ in the Hamiltonian family $H(\lambda)$, is a subset of the singular moduli $\mathcal{M}_s$, i.e., $\mathcal{M}_e \subseteq \mathcal{M}_s$. In most cases, where there is no extra symmetry, the two are equivalent. However, there are situations where $\mathcal{M}_e$ is a strict subset of $\mathcal{M}_s$, as we will explain in Subsec.~(\ref{sec:degenerate}). In the rest of the paper, we will not distinguish between the two unless an explicit statement is provided.

To construct the space $\mathcal{M}_s$ explicitly, one can compute the expectation values $\rho_i(\lambda) \coloneqq \Ave{\psi_e(\lambda)|H_i|\psi_e(\lambda)}$ for a given parameter set $\lambda$ and a corresponding eigenstate $\psi_e(\lambda)$. This corresponds to a point $(\rho_0(\lambda),\ldots,\rho_M(\lambda))$ in the projective space $\mathbb{RP}^M$. By continuously changing the parameter $\lambda$, we obtain a hypersurface that describes the singular moduli space $\mathcal{M}_s$, constructed by gluing together expectation values for all possible eigenstates. For $M>1$, the projective nature of the parameter set $\lambda$ implies that the lower and upper bounds are connected, resulting in a single closed boundary. In general, the moduli of the $i$-th and $(M-i)$-th eigenstates are connected for $0\leq i \leq M$.

This construction also implies that the boundary $\partial\mathcal{M} \subseteq \mathcal{M}_s$, which corresponds to the ground state (and the largest eigenvalue state) moduli, is a subset of $\mathcal{M}_s$. This is because the smallest and largest eigenvalues are the global lower and upper bounds of the energy functional (\ref{eq:E}). Figure~{\ref{fig:map}} illustrates the mapping $\rho$ from the Hilbert space $\mathcal{H}$ to the moduli space $\mathcal{M}$. The boundary $\partial\mathcal{M}$ provides a natural bound on expectation values, serving as a generalized uncertainty principle. 

The construction above suffers from two drawbacks. First, it requires solving the Schr\"odinger equation for all eigenstates of the entire Hamiltonian family, which is often impossible. Second, it does not provide an explicit equation for the hypersurface $\mathcal{M}_{e}$. To overcome these drawbacks and reveal the algebraic nature of $\mathcal{M}_{e}$, we start with the dual form of the Schr\"odinger equation, the cokernel condition~(\ref{eq:coker}). To have the existence of a non-trivial cokernel of the Jacobian $\J$, it is equivalent to require the vanishing of all $(M+1) \times (M+1)$ minors of $\mathcal{J}$, as expressed by the equation
\begin{equation}\label{eq:minor}
[\J]_{I,J} = 0,
\end{equation}
where $ [\J]_{I,J}$ represents the determinant of the submatrix of Jacobian $\J$ formed by selecting the rows of the index set $I$ and columns of the index set $J$, each with cardinality $M+1$. It is important to note that the minor condition~(\ref{eq:minor}) are polynomials in terms of $\psi$ alone and does not involve the parameter $\lambda$. The zero loci of these polynomials correspond to the eigenstate space $\mathcal{H}_e$, which consists of all eigenstates $\psi_e(\lambda)$ for the entire Hamiltonian family. The corresponding singular moduli $\mathcal{M}_s = \rho(\mathcal{H}_e)$ are given by the image of the mapping $\rho$ from the eigenstate space $\mathcal{H}_e$.

To determine the explicit equation for the image $\mathcal{M}_s$, we need to eliminate the variables $\psi$ and $\psi^*$ in Eq.~(\ref{eq:minor}) by substituting $\rho_i = \Ave{H_i}$ for $0\leq i\leq M$. This process can be carried out using ideal computation. Formally, we denote the ideal $\mathcal{I}$ over the polynomial ring $\mathbb{C}[\psi, \psi^*]$ as the one generated by the polynomials~(\ref{eq:minor}) \cite{wang1998jacobian}. The eigenstate space $\mathcal{H}_e$ is given by the corresponding homogeneous coordinate ring, i.e., $\mathcal{H}_e = \mathbb{C}[\psi, \psi^*]/\mathcal{I}$. We then extend the ring to the polynomial ring $\mathbb{C}[\psi, \psi^*, \rho]$ by including polynomials $\ave{H_i} - \rho_i = 0$, and compute the corresponding ideal. Finally, we project this extended ideal onto the subspace with $\rho$ alone. This computation is formally given by
\begin{equation} \label{eq:elimination}
\langle [\J]_{I,J}, \ave{H_i} - \rho_i \rangle \cap \mathbb{C}[\rho_i],
\end{equation}
where $\cap \mathbb{C}[\rho_i]$ reduces to the subideal in terms of $\rho$ and eliminates $\psi$ and $\psi^*$. In the end, we obtain a polynomial for the projective hypersurface $\mathcal{M}_s$ as
\begin{equation} \label{eq:fpr}
\fpr(\rho_0, \ldots, \rho_M) = 0.
\end{equation}
Practically, this elimination can be performed using Buchberger's algorithm with Gr\"obner basis, a nonlinear version of Gaussian elimination \cite{cox2013ideals}. We use the ``{\textsf GroebnerBasis}" function in {\textsf Mathematica} with the ``{\textsf EliminationOrder}" option to perform the elimination of Eq.~(\ref{eq:elimination}). Note that Eq.~(\ref{eq:fpr}) corresponds to the Zariski closure of $\mathcal{M}_s$, which may potentially include spurious branches. This is a necessary trade-off for achieving algebraic completeness. When $M+1 > 2N$, the minor condition~(\ref{eq:minor}) is trivially satisfied, and Eq.~(\ref{eq:elimination}) results in a set of algebraic equations that determine the variety $\mathcal{M} = \mathcal{M}_s$, as opposed to a single equation(\ref{eq:fpr}) for the singular hypersurface.

The polynomial~(\ref{eq:fpr}) is homogeneous, which means that we can replace $\rho_i = \ave{H_i}$ with $\Ave{H_i} = \frac{\ave{H_i}}{\langle\psi|\psi\rangle}$. This leads to $\fpr(\Ave{H_0},\ldots,\Ave{H_M}) = 0$ in terms of the expectation values $\Ave{H_i}$. Moreover, ${\Ave{H_i}}$ are linearly dependent since the identity operator belongs to the operator space $O$, i.e., $1 = \Ave{I} = \sum_i \alpha_i \Ave{H_i}$. Without loss of generality, we can choose the coordinate $H_0 = I$, which dehomogenizes Eq.~(\ref{eq:fpr}) into a hypersurface in the affine space $\mathbb{A}^M$, given by
\begin{equation} \label{eq:faf}
\faf(\Ave{H_1}, \ldots, \Ave{H_M}) \coloneqq \fpr(1, \Ave{H_1}, \ldots, \Ave{H_M}) = 0.
\end{equation}
This equation completely determines the relation between the expectation values $\Ave{H_i}$ for all eigenstates.

Below we demonstrate our construction for the expectation values of Pauli matrices $H_1 = \sigma_x$ and $H_2 = \sigma_y$.  We start with the Jacobian 
\begin{equation}
 \mathcal{J} = 
 \begin{bmatrix}
 \psi _1^* & \psi _2^* & \psi _1 & \psi _2 \\
 \psi _2^* & \psi _1^* & \psi _2 & \psi _1 \\
 i \psi _2^* & -i \psi _1^* & -i \psi _2 & i \psi _1 \\
\end{bmatrix},
\end{equation}
where we set $H_0 = I$, and $\psi = (\psi_1, \psi_2)$. We compute the minor condition (\ref{eq:minor}) explicitly for all $3\times 3$ minors, and although there are four equations, they reduce to a single equation, namely 
\begin{equation}
| \psi _2|^2-| \psi _1|^2 = 0.
\end{equation}
We extend the ring by including $\rho_0 \coloneqq \langle \psi | \psi \rangle$, $\rho_x \coloneqq \ave{\sigma_x}$, and $\rho_y \coloneqq \ave{\sigma_y}$. This gives us the following three additional equations:
\begin{equation}
\begin{split}
    |\psi_1|^2+| \psi_2|^2-\rho_0 &= 0 ,\\
   \psi_1 \psi_2^*+\psi_2 \psi_1^*-\rho_1 &= 0, \\
   i (\psi_1 \psi _2^*-\psi_2 \psi_1^*) -\rho_2 &= 0.
\end{split}
\end{equation}
We notice the identity $\rho_1^2 + \rho_2^2 - \rho_0^2 =  \left(| \psi _2|^2-| \psi _1|^2\right)^2 =  0$, which suggests $\fpr = \rho_1^2 + \rho_2^2 - \rho_0^2$, or equivalently, 
\begin{equation}\label{eq:sxy}
\faf = \Ave{\sigma_x}^2 + \Ave{\sigma_y}^2 - 1.
\end{equation}
As a result, the moduli $\mathcal{M}$ of these expectation values form a unit disk 
\begin{equation}\label{eq:sxy1}
\Ave{\sigma_x}^2 + \Ave{\sigma_y}^2 \leq 1,
\end{equation}
providing the uncertainty relation between $\sigma_x$ and $\sigma_y$. The boundary of this unit disk, $\partial\mathcal{M}$, is a unit circle given by Eq.~(\ref{eq:sxy}), corresponding to the ground states of the Hamiltonian family $H(\lambda) = \lambda_x \sigma_x + \lambda_y \sigma_y + \lambda_{-E} I$, as expected.

\subsection{Dual moduli space}\label{sec:polynomial}

We are not fully satisfied with the construction of $\mathcal{M}_s$ based on Eq.~(\ref{eq:elimination}). One of the main drawbacks is that it involves the wavefunction $\psi$ as an intermediate variable, which is then eliminated. In this subsection, we aim to find an alternative and more direct definition of $\mathcal{M}_s$, which will allow us to derive $\fpr$ and $\faf$ without going through the intermediate step of eliminating $\psi$.

Our observation starts with the simplest non-trivial case with $M=1$, where the operator space $O$ is generated by $\{I, H\}$, corresponding to the standard Schr\"odinger equation, $H \psi = E \psi$. In this case, the expectation value $\Ave{H}$ over an eigenvector gives rise to the corresponding energy $E$. Consequently, the singular moduli $\mathcal{M}_s$, in terms of $\Ave{H}$, are a set of isolated points determined by the roots of the characteristic polynomial of the Hamiltonian $H$, as
\begin{equation}\label{eq:cp}
\faf(\Ave{H}) = \det \left ( \Ave{H} - H \right).
\end{equation}
The moduli of the expectations $\mathcal{M}$ form a line segment bounded by the minimum and maximum eigenvalues. 

The above analysis suggests that the hypersurface equation~(\ref{eq:faf}) generalizes the characteristic polynomial of a single operator to an arbitrary set of operators. However, a direct generalization seems difficult. Therefore, we introduce first the dual singular moduli $\mathcal{M}_s^* \subseteq \mathbb{RP}^{M*}$, determined by
\begin{equation}\label{eq:fprd}
\fpr^*(\lambda_0,\ldots, \lambda_M) \coloneqq \det \left(\sum_{i=0}^M \lambda_i H_i\right) = 0,
\end{equation}
in terms of the dual variables $(\lambda_0, \ldots, \lambda_M)$. This equation gives rise to the constraint of the parameter set $\lambda$ for any eigenstates of the Schr\"odinger equation~(\ref{eq:schrodinger}). Additionally, the Schr\"odinger equation also requires
\begin{equation}\label{eq:dual}
\lambda_0 \rho_0 +\ldots +\lambda_M \rho_M = 0,
\end{equation}
for $(\rho_0,\ldots,\rho_M) \in \mathcal{M}_s$ and $(\lambda_0, \ldots, \lambda_M) \in \mathcal{M}_s^*$. 

Equation~(\ref{eq:dual}) implies the singular moduli $\mathcal{M}_s$ are projectively dual to  $\mathcal{M}_s^*$ \cite{tevelev2003projectively}. This duality means the points in the dual hypersurface $\mathcal{M}_s^*$ correspond to the tangent space of the original hypersurface $\mathcal{M}_s$, and vice versa. More specifically, we have the relation
\begin{equation} \label{eq:dfprd}
(\rho_0, \ldots, \rho_M) \sim (\partial_0 \fpr^*, \ldots, \partial_M \fpr^*),
\end{equation}
where $\sim$ denotes the proportionality of the left and right sides. One may view the projective dual as a geometric version of the Legendre transformation
\begin{equation} \label{eq:Legendre}
\tilde{f}_\mathrm{pr}(\rho ) \coloneqq \lambda(\rho)^t \rho - \fpr^*(\lambda(\rho)),
\end{equation}
where $\lambda(\rho)$ is the inverse of Eq.~(\ref{eq:dfprd}). In fact, for a point $(\rho_0,\ldots, \rho_M) \in \mathcal{M}_s$, Eqs.~(\ref{eq:dual}--\ref{eq:dfprd}) imply $\tilde{f}_\mathrm{pr}(\rho) = 0$. Therefore, the singular moduli $\mathcal{M}_s$ corresponds to the zero loci of Eq.~(\ref{eq:Legendre}). However, $\tilde{f}_\mathrm{pr}$ is generally non-algebraic. To obtain Zariski closure of $\tilde{f}_\mathrm{pr}$, that is, the  algebraic equation $\fpr$, one can use Eqs.~(\ref{eq:dual}--\ref{eq:dfprd}) to eliminate the dual variable $\lambda$. Buchberger's algorithm mentioned in the previous section can be used for this purpose, but more efficient algorithms exist for finding $\fpr$ based on its dual $\fpr^*$ \cite{bouziane2002computation}.

The dual singular moduli $\fpr^*$  provides valuable information about $\fpr$. For instance, the degree $d$ of polynomial $\fpr$ can be determined by the degree $d^* = N$ and singularity sets of $\fpr^*$. In the case of $M=2$, where both $\mathcal{M}_s$ and $\mathcal{M}_s^*$ are projective curves, Pl\"ucker formula \cite{hilton1920plane} provides a classical result
\begin{equation}\label{eq:plucker}
    d = d^*(d^*-1)-2\delta^*-3\kappa^*,
\end{equation}
where $\delta^*$ and $\kappa^*$ represent the number of ordinary double points and cusps of $\mathcal{M}_s^*$, respectively. Furthermore, the curves $\mathcal{M}_s$ and $\mathcal{M}_s^*$ share the same genus $g = \frac{1}{2}(d-1)(d-2)-\delta-\kappa$. For projective hypersurfaces with $M\geq 2$, the generalized Pl\"ucker formula has been found \cite{teissier1975diverses,parusinski1991multiplicity,ernstrom1997plucker}.

Below, we present an alternative method of deriving $\fpr$ and $\faf$. We start with the dual form of Eq.~(\ref{eq:dfprd})
\begin{equation}\label{eq:dfpr}
(\lambda_0, \ldots, \lambda_M) \sim (\partial_0 \fpr, \ldots, \partial_M \fpr).
\end{equation}
This equation indicates that the parameter set $(\lambda_0, \ldots, \lambda_M)$ corresponds to the normal vector of the hypersurface $\mathcal{M}_s$. Substituting Eq.~(\ref{eq:dfpr}) into Eq.~(\ref{eq:fprd}), we obtain:
\begin{equation} \label{eq:fg0}
h_\mathrm{pr} \fpr = \det \left ( \sum_{i=1}^{M} (\partial_i \fpr) H_i \right),
\end{equation}
where $h_\mathrm{pr}$ is the polynomial of the proportional factor. In other words, the right side must divide $\fpr$. Equation~(\ref{eq:fg0}) has multiple solutions, and we seek the polynomial solution $\fpr$ with the lowest degree $d$.

We can also convert Eq.~(\ref{eq:fg0}) into the affine version. We observe that $\partial_i \fpr(1,\ldots) = \partial_i \faf(\ldots)$ for $1\leq i \leq M$. Furthermore, substituting Eq.~(\ref{eq:dfpr}) into Eq.~(\ref{eq:dual}), we obtain $\partial_0 \fpr(1,\ldots) = - \sum_{i=1}^M \Ave{H_i} \partial_i \faf(\ldots)$. By substituting these equations into Eq.~(\ref{eq:fg0}), we obtain
\begin{equation} \label{eq:fg}
h \faf = \det \left ( \sum_{i=1}^{M} \partial_i \faf \left(\Ave{H_i}- H_i \right)\right),
\end{equation}
where $h(\ldots) \coloneqq -h_\mathrm{pr}(1,\ldots)$. For $M = 1$, we recover the characteristic polynomial~(\ref{eq:cp}), with $h = (\faf')^N$.

We will now use Eq.~(\ref{eq:fg}) to rederive $\faf(\Ave{\sigma_x}, \Ave{\sigma_y})$ of Eq.~(\ref{eq:sxy}). Due to rotational symmetry, we can make the ansatz that $\faf$ depends only on $\Ave{\sigma_x}^2+\Ave{\sigma_y}^2$. Substituting this ansatz into Eq.~(\ref{eq:fg}), we obtain
\begin{equation}
h\faf = (2\faf')^2 \det \left( \sum_{i=x,y} \Ave{\sigma_i} (\Ave{\sigma_i} - \sigma_i) \right ),
\end{equation}
where the determinant of the right side can be worked out explicitly as $(\Ave{\sigma_x}^2+\Ave{\sigma_y}^2)(\Ave{\sigma_x}^2+\Ave{\sigma_y}^2-1)$. Therefore, we have recovered Eq.~(\ref{eq:sxy}) with $h = 4(\Ave{\sigma_x}^2+\Ave{\sigma_y}^2)$.

\subsection{Heisenberg's uncertainty principle}\label{sec:HUP}

Our approach can be generalized to infinite-dimensional Hilbert spaces. For simplicity, we focus primarily on bound states. The expectation value moduli $\mathcal{M}$ and singular moduli $\mathcal{M}_s$ become real semianalytic sets and analytic varieties, respectively. It is worth noting that the operator space $O$ can also be infinite-dimensional, even uncountable. In the following, we will illustrate an example that is not only one of the most prominent examples but also demonstrates the usefulness of our approach: Heisenberg's uncertainty principle.

We consider the operator space $O$ generated by $\{I, x, x^2, p, p^2\}$. The corresponding moduli space $\mathcal{M}$ is a four-dimensional space consisting of the expectation values $\Ave{x^2}$, $\Ave{p^2}$, $\Ave{x}$, and $\Ave{p}$. To find the corresponding $\faf$, we make the ansatz that $\faf$ depends only on $\Delta x^2 \Delta p^2$, where $\Delta x^2 \coloneqq \langle x^2 \rangle - \langle x \rangle^2 $ and $\Delta p^2 \coloneqq \langle p^2 \rangle - \langle p \rangle^2$. Substituting this ansatz into Eq.~(\ref{eq:fg}), we obtain
\begin{equation}\label{eq:fHUP}
h\faf = \det \left( 2\Delta x^2 \Delta p^2 - \Delta p^2 (x-\Ave{x})^2 - \Delta x^2 (p-\Ave{p})^2 \right) \faf'.
\end{equation}
To evaluate the functional determinant, we first diagonalize the operator $H= \Delta p^2 (x-\Ave{x})^2 + \Delta x^2 (p-\Ave{p})^2$. This can be done by introducing the annihilation operator $a = (2\hbar\alpha \beta)^{-1/2} \left(\alpha (x-\Ave{x}) + i \beta (p-\Ave{p})\right)$, which satisfies the commutation relation $[a,a^\dag] = 1$. Here, the real parameters $\alpha$ and $\beta$ are determined by matching $H$ to the diagonal form $2\hbar \alpha \beta (a^\dag a + 1/2) = \alpha^2 (x-\Ave{x})^2 + \beta^2 (p-\Ave{p})^2$, which indicates that $\alpha^2 = \Delta x^2$ and $\beta^2 = \Delta p^2$. The eigenvalues can then be computed as $E_n = 2\hbar \alpha \beta (n+1/2) = 2\sqrt{\Delta x^2 \Delta p^2} \hbar (n+1/2)$. Therefore, the zeros of Eq.~(\ref{eq:fHUP}) are $\Delta x^2 \Delta p^2 = (\hbar (n+1/2))^2 $.  To ensure that $\faf$ is well-defined, an appropriate regularization is required due to the infinite-dimensional Hilbert space. This regularization can be absorbed by the factor $h$ and leads to a convergent determinant $\faf = \prod_{n = 0}^\infty (1 - \Delta x^2 \Delta p^2 / ((n + 1/2)\hbar)^2)$, or equivalently
 \begin{equation}\label{eq:HUP1}
    \faf = \cos \left( \frac{\pi \sqrt{\Delta x^2 \Delta p^2}}{\hbar}\right).
 \end{equation}
This suggests that the singular moduli $\mathcal{M}e$ is an analytic variety. The smallest root of Eq.(\ref{eq:HUP1}) gives rise to the boundary $\partial \mathcal{M}$, which leads to Heisenberg's uncertainty principle \cite{Heisenberg1927, kennard1927quantenmechanik, weyl1950theory}
\begin{equation}\label{eq:HUP}
\Delta x^2 \Delta p^2 \geq \hbar^2/4,
\end{equation}
for the expectation moduli $\mathcal{M}$, as shown in Fig.\ref{fig:HO}. The uncertainty principle~(\ref{eq:HUP}) provides a nonlinear bound on the expectation values of physical observables that involve quadratic forms of position and momentum operators. The singular moduli $\mathcal{M}_s$ correspond to the expectation values of the eigenstates of the harmonic oscillator $ H(\lambda) = \lambda_{p^2} p^2 + \lambda_{x^2} x^2 + \lambda_{p} p + \lambda_{x} x + \lambda_{-E} I$, and this bound is saturated by the ground state, as expected. It is worth noting that Equation~(\ref{eq:HUP1}) can also be derived by first calculating the dual $\fpr^*$ using the functional determinant of $H(\lambda)$ in Eq.~(\ref{eq:fprd}), and then applying the Legendre transform~(\ref{eq:Legendre}).

\begin{figure}
 \includegraphics[width=1\linewidth]{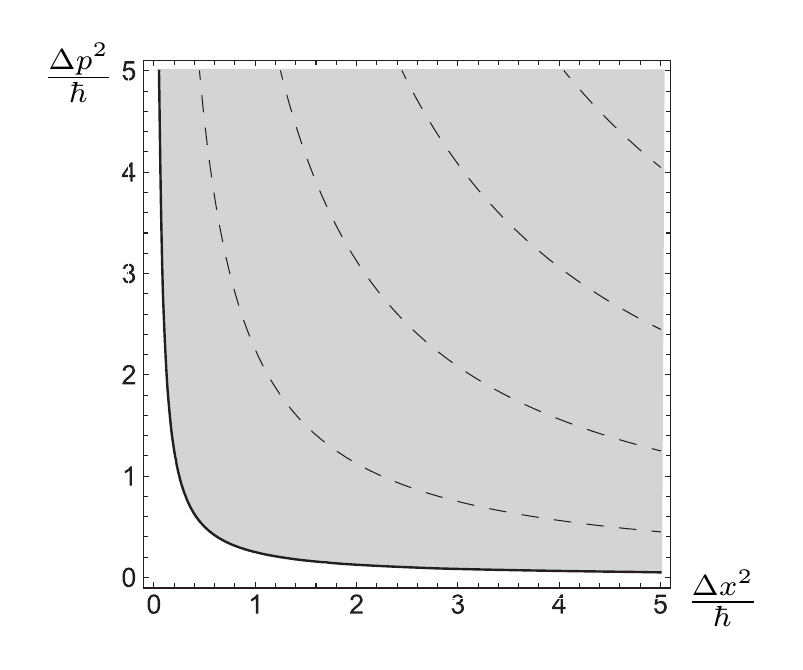}
 \caption{The moduli space $\mathcal{M}$ for operators $\Ave{x}$, $\Ave{x^2}$, $\Ave{p}$ and $\Ave{p^2}$ is represented by the grey domain and bounded by Heisenberg's uncertainty principle. The solid curve represents the ground state moduli, saturating the uncertainty principle. Dashed curves correspond to excited states. }
 \label{fig:HO}
\end{figure}

We next include the operator $\frac{1}{2}\{x,p\} = \frac{1}{2}(x p + p x)$ to the generators of the operator space $O$. We adopt a similar approach as above and propose that $\faf$ is a function of $\Delta x^2 \Delta p^2 - (\Delta xy)^2$, where $\Delta xy \coloneqq \frac{1}{2}\Ave{{x,p}} - \Ave{x} \Ave{p}$. Substituting this into Eq.~(\ref{eq:fg}), we obtain
\begin{equation}\label{eq:SRf}
        h\faf = \det \faf' ( 2(\Delta x^2 \Delta p^2 - (\Delta xy)^2)- H),
\end{equation}
where $H \coloneqq  \Delta p^2  (x-\Ave{x})^2 + \Delta x^2  (p-\Ave{p})^2 - \Delta xy \{x-\Ave{x},p -  \Ave{p}\}$. To diagonalize the operator $H$, we introduce the annihilation operator $a = (2\hbar \Re(\alpha  \beta^*))^{-1/2} \left(\alpha (x-\Ave{x}) + i \beta (p-\Ave{p})\right)$, where the parameters $\alpha$ and $\beta$ are complex numbers. Matching $H$ to the diagonal form $2\hbar \Re(\alpha  \beta^*) (a^\dag a + 1/2)$ yields $|\alpha|^2 = \Delta x^2$, $|\beta|^2 = \Delta p^2$, $\Im (\alpha \beta^*) = \Delta xy $. Note that $\Re(\alpha \beta^*)^2 = |\alpha|^2|\beta|^2 - \Im(\alpha \beta^*)^2 = \Delta x^2 \Delta y^2 - (\Delta xy)^2$. The eigenvalues can be computed as $E_n = 2\hbar \Re(\alpha \beta^*) (n+1/2) = 2\sqrt{\Delta x^2 \Delta y^2 - (\Delta xy)^2} \hbar (n+1/2)$. Therefore, the zeros of Eq.~(\ref{eq:fHUP}) are given by $\Delta x^2 \Delta p^2 - (\Delta xy)^2 = (\hbar (n+1/2))^2 $, leading
 \begin{equation}\label{eq:SRUP}
    \faf = \cos \left( \frac{\pi \sqrt{\Delta x^2 \Delta p^2 - (\Delta xy)^2}}{\hbar}\right).
 \end{equation}
The lowest root of Eq.~(\ref{eq:SRUP}) gives rise to the Schr\"odinger-Robertson uncertainty relation \cite{robertson1929uncertainty,schrodinger1930sitzungsberichte}. 
 \begin{equation}
     \Delta x^2 \Delta p^2 - (\Delta xy)^2 \geq \hbar^2/4.
 \end{equation}

One may apply our analysis to higher-order operators, such as $x^4$ and $p^4$, by including them in the operator space $O$, thereby obtaining an uncertainty relation beyond the quadratic level. For example, applying the same argument used previously, we obtain the following inequality
\begin{equation}
    \Ave{x^{2n}}\Ave{p^{2n}} \geq \eta_{n}^2(\hbar^{2n} /4).
\end{equation}
Here, $n$ is a positive integer, and $\eta_{n}$ represents the lowest eigenvalue of the operator $(-1)^n \frac{d^{2n}}{d x^{2n}} + x^{2n}$.  Note that $\eta_1 = 1$ corresponds to Heisenberg's uncertainty principle~(\ref{eq:HUP}). One can show that $2^{1-n} \leq \eta_{n} \leq 2^{1-n} (2n-1)!!$. The lower bound results from H\"older's inequality, which yields $\Ave{x^{2n}} \geq \Ave{x^{2}}^n$ and $\Ave{p^{2n}} \geq \Ave{p^{2}}^n$, in conjunction with Eq.~(\ref{eq:HUP}). The upper bound is estimated using the trial wavefunction $\psi = \frac{1}{\sqrt{\pi}}e^{-x^2/2}$. Numerical calculations reveal that $\eta_{2} \approx 1.40$ and $\eta_{3} \approx 2.95$, which are closer to the upper bound values. The exact values of $\eta_n$ and their asymptotic behavior merit further investigation in future studies.

In general, finding an explicit formula by solving Eq.~(\ref{eq:fg}) may be challenging. Nevertheless, our approach offers a general framework for generalizing Heisenberg's uncertainty principle for any set of operators. In the case of continuous spectra, the set of singular moduli $\mathcal{M}_s$ is dense and equivalent to $\mathcal{M}$. However, it still makes sense to discuss the boundary geometry $\partial \mathcal{M}$, which corresponds to ground states. 

\subsection{Degeneration and integrable system}\label{sec:degenerate}

In this subsection, we briefly discuss the case of degeneracy. When the operator space $O$ degenerates, it can be extended by central extension, where the center $\mathfrak{z}(O)$ commutes with all Hamiltonians. This central extension leads to a non-trivial symmetry group $Z \coloneqq \exp(i \mathfrak{z}(O)) \subset U(N)$, implying that the cokernel of the eigenstate Jacobian ${\mathcal{J}}(\psi_e)$ typically has a larger dimension than one, which is stronger than the cokernel condition~(\ref{eq:coker0}). 

To understand this better, in the presence of extra symmetry, the Hilbert space $\mathcal{H}$ decomposes as a direct sum $\mathcal{H} = \bigoplus_\alpha \mathcal{H}^{(\alpha)}$, where $\mathcal{H}^{(\alpha)}$ are the irreducible Hilbert spaces. For an eigenstate $\psi_e \in \mathcal{H}^{(\alpha)}$, the corresponding irreducible representation of the Hamiltonian $H^{(\alpha)}(\lambda)$ has a smaller dimension $N_\alpha \coloneqq \dim \mathcal{H}^{(\alpha)}$, which often have null vectors
\begin{equation}
\sum_{i=1}^M \lambda^{(\alpha)}_i \mathbf{H}^{(\alpha)}_i = 0.
\end{equation}
Here, $\lambda^{(\alpha)}$ represents the cokernel of the irreducible Hilbert space $ \mathcal{H}^{(\alpha)}$, which reflects the global symmetry and is independent of the choice of state. Therefore, the solution $\lambda$ of Eq.~(\ref{eq:coker}) for a given eigenstate $\psi_e$ admits a larger solution space generated by both $\lambda$ and $\lambda^{(\alpha)}$, leading to $\dim \mathrm{coker} \mathcal{J}(\psi_e) = 1+ \dim \lambda^{(\alpha)}$.

In this case, the eigenstates moduli $\mathcal{M}_e$ is a strict subset of the singular moduli $\mathcal{M}_s$, i.e., $\mathcal{M}_e \subsetneq \mathcal{M}_s$. Geometrically, this implies a stratification of the singular moduli  $\mathcal{M}_s$, where its $r$-codimensional strata $\mathcal{S}_r$ corresponds to the Jacobian with cokernel dimension $r$. In particular, the eigenstate moduli $\mathcal{M}_e$ corresponds to strata with a smaller dimension. 

We now consider the extreme case where all operators commute, i.e., $[ H(\lambda), H(\lambda')] = 0$. In this case, the dimension of the center of the operator space $\mathfrak{z}(O)$ is $M$, and consequently, the eigenstate moduli $\mathcal{M}e$ reduces to a set of isolated points. 
To gain better insight into this result, we observe that it is possible to simultaneously diagonalize all operators with the same set of eigenstates ${\psi_e}$. As a result, the dual singular moduli $\mathcal{M}_s^*$ in Eq. (\ref{eq:fprd}) factorizes as
\begin{equation}\label{eq:dualfactor}
\fpr^*(\lambda_0,\ldots, \lambda_M) = \prod_{e} \left(\sum_{i=0}^M \lambda_i h_i^e\right),
\end{equation}
where $h_i^e$ is the eigenvalue of operator $H_i$ for the eigenstate $\psi_e$. 

Equation~(\ref{eq:dualfactor}) implies that the dual singular moduli $\mathcal{M}^*$ is a polytope enclosed by the set of faces $\sum_{i=0}^M \lambda_i h_i^e = 0$. The expectation value moduli $\mathcal{M}$ is its dual polytope, whose boundary consists of a set of dual faces, each corresponding to a vertex of $\mathcal{M}^*$. Conversely, the vertices of $\mathcal{M}$ correspond to the faces in $\mathcal{M}^*$ that satisfy $\sum_{i=0}^M \lambda_i h_i^e = 0$,  i.e., an eigenstate in the moduli $\mathcal{M}_e$.

When a system is completely integrable, we can identify a maximal set of $N$ commuting operators. In this case, the face equations are obtained by taking the inverse of the coefficient matrix $[h_i^e]$, and a detailed construction is described in Reference~\cite{song2023zero}. For $1+1$D integrable systems, the Bethe ansatz provides a way to determine these equations \cite{bethe1931theorie,yang1968s,baxter1972partition,takhtadzhan1979quantum,jimbo1990yang,gromov2014quantum}. In particular, Eq.~(\ref{eq:faf}) factorizes into the product of the expectation values of the transfer matrix, evaluated at its polynomial roots.

\subsection{Relations between expectation moduli}\label{sec:cat}

\begin{figure*}
 \includegraphics[width=1\linewidth]{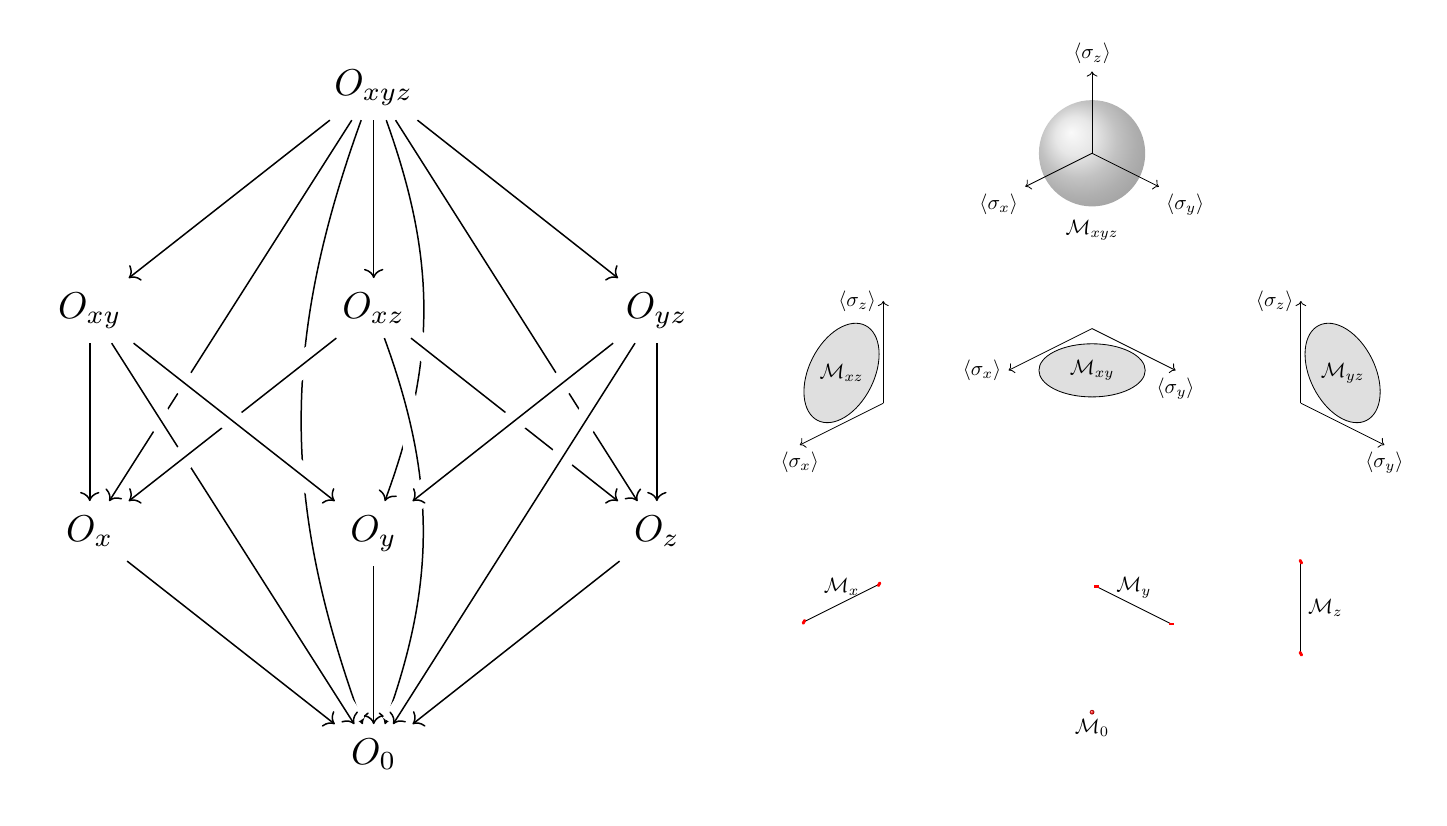}
 \caption{Left: Eight operator spaces are shown for the two-dimensional complex Hilbert space, where the projection from space $O$ to $O'$ is represented by arrows. Right: The corresponding moduli space $\mathcal{M}_{i} = \mathcal{M}(O_{i})$ is depicted, with the same projection relation. The arrow is hidden for ease of readability.}
 \label{fig:su2}
\end{figure*}

Up to this point, our focus has been solely on the moduli space $\mathcal{M}(O)$ for a fixed operator space $O \subseteq \Ls$. In this subsection, we investigate the relationships between all operator spaces and their moduli spaces.  Given two operator spaces $O$ and $O'$, if $O' \subseteq O$ is a subspace of $O$, there is a natural projection mapping $\pi_0$ from $O$ to $O'$. For example, consider a two-dimensional complex Hilbert space with the following eight operator spaces: $O_{xyz} \coloneqq \Ls = \sp{\sigma_x, \sigma_y, \sigma_z, I} $, $O_{xy} \coloneqq \sp{\sigma_x, \sigma_y, I}$, $O_{xz} \coloneqq \sp{\sigma_x, \sigma_z, I}$, $O_{yz} \coloneqq \sp{\sigma_y, \sigma_z, I}$, $O_x \coloneqq \sp{\sigma_x, I}$, $O_y \coloneqq \sp{\sigma_y, I}$, $O_z \coloneqq \sp{\sigma_z, I}$, and $O_0 \coloneqq \sp{I}$. The possible projections between these subspaces are depicted by arrows in the left panel of Fig.~\ref{fig:su2}.

We can lift a projection mapping $\pi_0$ from an operator space $O$ to its subspace $O'$ to a projection from the moduli space $\mathcal{M}(O)$ to $\mathcal{M}(O')$. This can be illustrated by the following commutative diagram: 
\begin{center}
\begin{tikzcd}
O \arrow[r] \arrow[d, "\pi_0"]
& \mathcal{M}(O) \arrow[d, "\pi"] \\
O' \arrow[r]
& \mathcal{M}(O')
\end{tikzcd} %
\end{center}
Here, $\pi$ denotes the pullback induced by the projection map $\pi_0: O' \rightarrow O$, which maps an element of $\mathcal{M}(O)$ to its corresponding element in $\mathcal{M}(O')$. The commutativity arises from the fact the mapping $\rho$ is linear in terms of operators, i.e., $\rho(a H_1+b H_2) = a \rho(H_1) + b \rho(H_2)$, and $\pi_0$ is a linear projection. This means that if two operator spaces $O$ and $O'$ are related by $O' \subseteq O$, then there exists a well-defined linear projection $\pi$ between their moduli spaces $\mathcal{M}(O)$ and $\mathcal{M}(O')$. 

As any operator space $O$ is a subspace of the real vector space consisting of all self-adjoint operators $\Ls$, the corresponding moduli $\mathcal{M}(O)$ can be constructed via the projection $\pi$ from the moduli $\mathcal{M}(\Ls)$. For a $N$-dimensional complex Hilbert space $\mathcal{H}$, any operator $A \in \Ls$ can be decomposed as $A = \sum_{i,j} A_{ij} e^{(ij)}$ in terms of $N^2$ basis $e^{(ij)}$, where $e^{(ij)}$ is the matrix with a unit in the $(i,j)$ position and zeros elsewhere. It is worth noting that the expectation value $\Ave{e^{(ij)}} = \psi_i^* \psi_j $ satisfies a set of quadratic equations known as Veronese embedding \cite{hartshorne2013algebraic}, given by
\begin{equation}\label{eq:vernose}
     \Ave{e^{(ij)}} \Ave{e^{(kl)}} =  \Ave{e^{(il)}}  \Ave{e^{(kj)}},
\end{equation}
for any set of indices $1\leq i,j,k,l\leq N$, which determines completely the moduli $\mathcal{M}(\Ls)$ as a projective variety. 
Introducing the operator matrix $\mathbf{e} = [e^{(ij)}]$ with $e^{(ij)}$ as its element at position $(i,j)$, Equation~(\ref{eq:vernose}) is equivalent to the vanishing of all $2\times 2$ minors of the density matrix $\Ave{\mathbf{e}} =  |\psi\rangle\langle \psi|$, since it has rank one.

Consider the previous example with $N=2$. In this case, the operator matrix is given by $\mathbf{e} = \frac{1}{2} \begin{bmatrix}
    I_2 + \sigma_z & \sigma_x + i \sigma_y \\
    \sigma_x - i \sigma_y  & I_2 - \sigma_z
\end{bmatrix}$. To make the $2\times 2$ minors of the density matrix $\Ave{\mathbf{e}}$ vanish, we require $\det \Ave{\mathbf{e}} = 0 $. This leads to 
\begin{equation}\label{eq:su2}
\Ave{\sigma_x}^2+ \Ave{\sigma_y}^2+ \Ave{\sigma_x}^2 = 1,
\end{equation}
which suggests that $\mathcal{M}(\Ls)$ is a three-dimensional unit sphere. The expectation moduli of any operator subspace can be constructed directly from the projection of Eq.~(\ref{eq:su2}), as shown in the right panel of Fig.~\ref{fig:su2}. For instance, for the subspace $O_{xy}$ generated by $\sigma_x$ and $\sigma_y$, the projection leads to a unit disk, recovering our earlier result of Eq.~(\ref{eq:sxy}). 

To gain a deeper understanding of the moduli space for many-body systems, below we consider a scenario where the Hilbert space $\mathcal{H}_{AB}$ is the tensor product of two Hilbert spaces, $\mathcal{H}_A$ and $\mathcal{H}_B$, i.e., $\mathcal{H}_{AB} = \mathcal{H}_A \otimes \mathcal{H}_B$. This defines a tensor product of the moduli spaces $\mathcal{M}(\mathcal{L}_s(\mathcal{H}_A))\otimes\mathcal{M}(\mathcal{L}_s(\mathcal{H}_B)) \coloneqq \mathcal{M}(\mathcal{L}_s(\mathcal{H}_{AB}))$. This tensor product can be induced through the tensor product of the density matrix, 
\begin{equation} \label{eq:eAB}
    \Ave{\mathbf{e}_{AB}}=\Ave{\mathbf{e}_A\otimes\mathbf{e}_B},
\end{equation} 
which defines the moduli $\mathcal{M}(\mathcal{L}_s(\mathcal{H}_{AB}))$ via Eq.~(\ref{eq:vernose}), without requiring information about the original Hilbert spaces $\mathcal{H}_A$ and $\mathcal{H}_B$.

Now let's consider the tensor product of subspaces $\mathcal{M}(O_A) \otimes \mathcal{M}(O_B)\coloneqq \mathcal{M}(O_A\otimes O_B)$, where $O_A \subseteq \mathcal{L}_s(\mathcal{H}_A)$ and $O_B \subseteq \mathcal{L}_s(\mathcal{H}_B)$. As we discussed earlier, there exist linear projections $\pi_a: \mathcal{M}(\mathcal{L}_s(\mathcal{H}_A)) \longrightarrow \mathcal{M}(O_A)$ and $\pi_b: \mathcal{M}(\mathcal{L}_s(\mathcal{H}_B)) \longrightarrow \mathcal{M}(O_B)$. Their tensor product $\pi_a \otimes \pi_b: \mathcal{M}(\mathcal{L}_s(\mathcal{H}_{AB})) \longrightarrow \mathcal{M}(O_A\otimes O_B)$ induces a linear projection from $\mathcal{M}(\mathcal{L}_s(\mathcal{H}_{AB}))$, which uniquely determines the tensor-producted moduli $\mathcal{M}(O_A) \otimes \mathcal{M}(O_B)$.

This construction enables us to study quantum entanglement using only the moduli spaces and their tensor products. Notably, some of the $2\times 2$ minors in Eq.~(\ref{eq:eAB}) involve elements from both systems $A$ and $B$, leading to nontrivial constraints on the tensor-producted moduli space. This is distinct from the expectation moduli of separable states, as we will demonstrate in Subsec.~\ref{sec:bell}.

We can formally express our analysis above using the language of category theory. Given a Hilbert space $\mathcal{H}$ and the real vector space $\Ls$ of self-adjoint operators, we introduce the category of operator spaces, denoted $\textbf{Op}$. The objects in this category are subspaces $O \subseteq \Ls$, including the identity operator $I \in O$. The morphisms in $\textbf{Op}$ are inclusions between subspaces. All relevant physical information is contained within the category $\textbf{Op}$, which has an initial object $\mathcal{I}_{Op} = \Ls$ and a terminal object $\mathcal{T}_{Op} = \{ I \}$. We can also introduce the category of expectation value moduli, denoted $\textbf{Ev}$. The objects in this category, $\mathcal{M}(O)$, represent moduli of expectation values of $O$. The morphisms in $\textbf{Ev}$ are linear projections between objects. There exists a natural functor from the operator category $\textbf{Op}$ to the expectation value category $\textbf{Ev}$. 

As we demonstrated above, constructing $\textbf{Ev}$ does not necessarily require introducing the Hilbert space $\mathcal{H}$. To see this, we first need to construct the initial object $\mathcal{I}_{Ev} = \mathcal{M}(\Ls)$, which can be defined by Eq.~(\ref{eq:vernose}) for finite-dimensional cases. For infinite-dimensional cases, one may construct $\mathcal{I}_{Ev}$ using a similar argument as the finite-dimensional one. An object $E_i$ in $\textbf{Ev}$ is determined by the linear projection $\pi_i$ from the initial object $\mathcal{I}_{Ev}$ to it. All physical measurements are directly read from $\textbf{Ev}$, either from the geometry of an object (ground state expectations) or the morphism from $\mathcal{I}_{Ev}$ to it (eigenstate expectations). Thus, the category $\textbf{Ev}$ contains all time-independent physics without involving the Hilbert space $\mathcal{H}$, potentially providing a new time-independent quantum formulation. Building on our earlier discussion, we can further introduce the tensor product of two moduli categories, which yields a monoidal category of moduli categories. This will be useful for future investigations of quantum entanglement.

\subsection{Classical moduli space}\label{sec:classical}

In this subsection, we will discuss the moduli space $\mathcal{M}_\mathrm{cl}$ as the classical counterpart of the proposed quantum moduli space $\mathcal{M}$. We start with a $2n$-dimensional symplectic space $\mathbb{R}_{\mathbf{q},\mathbf{p}}^{2n}$ equipped with the canonical symplectic form $\omega = dq_1\wedge dp_1 + \ldots + dq_n \wedge dp_n$, and consider a $M$-dimensional real vector space $O_\mathrm{cl}$ of classical physical observables, generated by a set of basis $\{H_1^\mathrm{(cl)}(\mathbf{q},\mathbf{p}), \ldots, H_M^\mathrm{(cl)}(\mathbf{q},\mathbf{p})\}$. To define the classical moduli space $\mathcal{M}_\mathrm{cl}$, we introduce a mapping $\rho_\mathrm{cl}$ from the orbit space $\Gamma$ of  $\mathbb{R}_{\mathbf{q},\mathbf{p}}^{2n}$ to the $M$-dimensional real affine space $\mathbb{A}^M$, given by
\begin{equation}
\rho_\mathrm{cl}(\gamma) \coloneqq \left( \Ave{H_1^\mathrm{(cl)}}_\gamma , \ldots, \Ave{H_{M}^\mathrm{(cl)}}_\gamma \right),
\label{eq:rhocl}
\end{equation}
where $\gamma = (\mathbf{q} (t),\mathbf{p}(t))$ denotes an orbit, and 
\begin{equation}
\Ave{H_i^\mathrm{(cl)}}_\gamma \coloneqq \frac{1}{T_\gamma} \int_\gamma H_i^\mathrm{(cl)}(\mathbf{q},\mathbf{p}) dt,
\end{equation}
represents the classical expectation value over the orbit $\gamma$, where $T_\gamma$ is the period for periodic orbits, or $T_\gamma \to \infty$ for quasiperiodic or chaotic orbits. We define the classical moduli space $\mathcal{M}_\mathrm{cl}$ as the image of the mapping $\rho_\mathrm{cl}$.

The geometry of $\mathcal{M}_\mathrm{cl}$ itself provides little information, as moduli $\mathcal{M}_\mathrm{cl}$ are typically simple. For instance, in the case of quadratic forms $x$, $x^2$, $p$, and $p^2$, it reduces to $\Delta x_\mathrm{cl}^2 \geq 0$ and $\Delta p_\mathrm{cl}^2 \geq 0$. By contrast, the quantum moduli space $\mathcal{M}$ given by the uncertainty principle~(\ref{eq:HUP}) provides a tighter and non-trivial bound. 

On the other hand, the singular set of the mapping $\rho_\mathrm{cl}$ offers more interesting information, leading to the so-called reciprocal Maupertuis principle \cite{gray1996four}, given by 
\begin{equation} \label{eq:clvar}
    \delta \frac{1}{T_\gamma} \int_{\gamma} H^\mathrm{(cl)}(\mathbf{q},\mathbf{p}) dt = 0,
\end{equation}
where $H^\mathrm{(cl)}(\mathbf{q},\mathbf{p}) = \sum_{i=1}^M \lambda_i H_i^\mathrm{(cl)}(\mathbf{q},\mathbf{p})$. Equation~(\ref{eq:clvar}) represents the classical limit of the quantum variational principle,
\begin{equation}
    \delta \frac{\ave{H}}{\langle\psi|\psi\rangle} = 0.
\end{equation}
Solving Eq.~(\ref{eq:clvar}) yields the classical equation of motion. Therefore, similar to the quantum version, the singular set of the mapping $\rho_\mathrm{cl}$ corresponds to the orbits $\gamma$ under the classical Hamiltonian flow. 

Our discussion above suggests a potential ``quantization" from the classical moduli space to the quantum ones. However, since the classical energy spectrum is dense, the corresponding singular set has the same geometry as $\mathcal{M}_\mathrm{cl}$. To obtain the appropriate classical counterpart of singular moduli $\mathcal{M}_s$, we need to employ the semiclassical quantization. One way to accomplish this is by using the saddle point approximation of the path integral employed by Gutzwiller to derive his famous semiclassical trace formula \cite{gutzwiller1971periodic}. As a result, the semiclassical approximation of the quantum spectral determinant $\det(E-H)$ leads to the Voros-Gutzwiller zeta function \cite{voros1988unstable},
\begin{equation}\label{eq:sczeta}
\zeta_\mathrm{sc}(E;H^\mathrm{(cl)}) = \exp\left( -\sum_{p} \sum_{n=1}^\infty \frac{1}{n} \frac{e^{in(E T_p - \nu_p \pi /2)}}{|\Lambda_p|^{n/2}(1-\Lambda_p^{-n})}\right),
\end{equation}
summing over all primary periodic orbits $p$, with $\nu_p$ and $\Gamma_p$ denoting the corresponding Maslov index and the dominant eigenvalue of the stability matrix, respectively. By combining Eq.(\ref{eq:fprd}), we obtain the semiclassical equation for the dual singular moduli $\mathcal{M}_s^\mathrm{(cl)*}$, 
\begin{equation}
f^*_\mathrm{sc}(\lambda_0,\ldots, \lambda_M) \coloneqq \zeta_\mathrm{sc}\left(-\lambda_0;\sum_{i=1}^M \lambda_i H_i^\mathrm{(cl)}\right), 
\end{equation}
where we set $H_0 = I$ and $\lambda_0 = -E$.  The Legendre transformation~(\ref{eq:Legendre}) can then be used to determine the semiclassical equation $f_\mathrm{sc}$ for the singular moduli $\mathcal{M}_s^\mathrm{(cl)}$. By continuously varying the classical Hamiltonian $H^\mathrm{(cl)}$ with the parameter set $(\lambda_1, \ldots, \lambda_M)$, periodic orbits undergo continuous deformation, potentially resulting in rich geometry such as bifurcations. Exploring the connection between this geometry and the semiclassical singular moduli $\mathcal{M}_s^\mathrm{(cl)}$ will be the focus of future investigations.

\subsection{Finite temperature} \label{sec:FT}

In this subsection, we consider the finite temperature case. We begin by setting $H_0 = I$, $\lambda_0 = -\beta E$ and defining the partition function as
\begin{equation}
\mathcal{Z}(\lambda_1,\ldots, \lambda_M) \coloneqq \mathrm{tr} \exp\left(-\sum_{i=1}^M \lambda_i H_i\right).
\end{equation}
The free energy, given by
\begin{equation}\label{eq:W}
W(\lambda_1, \ldots, \lambda_M) \coloneqq -\ln \mathcal{Z},
\end{equation}
plays a role similar to the dual $\fpr^*$ defined in Eq.~(\ref{eq:fprd}). In fact, they are closely related since the partition function $Z$ is the Laplace transform of the density of states $g(E) = \frac{1}{\pi}\Im \partial_E \ln \fpr^*(-E-i0^-,\ldots)$.

We introduce the thermal expectation
\begin{equation}\label{eq:HW}
\overline{H_i} = \partial_i W(\lambda),
\end{equation}
and the Legendre transformation in analogy to Eq.~(\ref{eq:Legendre}), given by
\begin{equation}\label{eq:Phi}
\Phi(\overline{H}) \coloneqq W(\lambda(\overline{H})) - \lambda(\overline{H})^t \overline{H},
\end{equation}
where $\lambda(\overline{H})$ is the inverse of Eq.~(\ref{eq:HW}). The dual variable is now given by
\begin{equation}\label{eq:PhiH}
\lambda = -\partial_i \Phi(\overline H).
\end{equation}

We can generalize this approach to quantum field theory via path integrals, where the partition function is defined as
\begin{equation}
\mathcal{Z}(\lambda_1,\ldots, \lambda_M) \coloneqq \int \exp\left(- S[\phi|\lambda] \right) D[\phi],
\end{equation}
with the action functional $S[\phi|\lambda]$ a linear superposition in terms of the parameter set $\lambda$. Equations (\ref{eq:W}--\ref{eq:PhiH}) follow sequentially. One of the most notable examples of this framework is the Luttinger–Ward functional \cite{luttinger1960ground}, where 
\begin{equation}
    S = S_0 + \int \phi_i(x)^\dag \lambda(x,x') \phi_j(x') d^4x d^4x'.
\end{equation}
Here the parameter set $\lambda_0 = 1$, and $\lambda(x,x')$ is now an uncountable set. The corresponding thermal expectation~(\ref{eq:HW}) is
\begin{equation}
    \frac{\delta{W}[\lambda]}{\delta \lambda} = \overline{\phi(x)^\dag \phi(x')} \equiv G(x,x'),
\end{equation}
which is the Green's function. The functional $\Phi$ defined in Eq.~(\ref{eq:Phi}) is known as the Baym–Kadanoff functional \cite{baym1961conservation}. By subtracting the non-interacting term, we obtain the Luttinger–Ward functional.

\section{Applications}\label{application}

\subsection{Quantum nonlocality}\label{sec:bell}

In this subsection, we utilize our theory to investigate quantum nonlocality, starting with two possible measurements performed by Alice ($A_0$ and $A_1$) and Bob ($B_0$ and $B_1$) standing in widely separated locations.  For all subsequent discussions, we set $A_0 = \sigma_z^a$, $A_1 = \sigma_x^a$, $B_0 = -\frac{1}{\sqrt{2}}(\sigma_x^b+\sigma_z^b)$, and $B_1 = \frac{1}{\sqrt{2}}(\sigma_x^b - \sigma_z^b)$. The celebrated Bell inequality~\cite{bell1964einstein},
\begin{equation}\label{eq:bell}
\left|\Ave{A_0 B_0}+\Ave{A_0 B_1}+\Ave{A_1 B_0}-\Ave{A_1 B_1}\right| \leq 2,
\end{equation}
sets an upper limit for local hidden-variable theory. However, quantum mechanics can violate Bell inequalities, and the maximum violation is determined by the Tsirelson bound \cite{cirel1980quantum},
\begin{equation}\label{eq:tsirelson}
\left| \Ave{A_0 B_0}+\Ave{A_0 B_1}+\Ave{A_1 B_0}-\Ave{A_1 B_1} \right| \leq 2\sqrt{2}.
\end{equation}
Notably, the inequality~(\ref{eq:tsirelson}) saturates when $\left(\Ave{A_0 B_0},\Ave{A_0 B_1},\Ave{A_1 B_0},\Ave{A_1 B_1}\right) = \frac{1}{\sqrt{2}}(1,1,1,-1)$, corresponding to the Bell state $\psi = \frac{1}{\sqrt{2}}(|0\rangle\otimes|1\rangle - |1\rangle\otimes|0\rangle)$.

The inequality~(\ref{eq:tsirelson}), however, provides only an upper bound, i.e., not all expectation values that satisfy this equation are allowed quantum mechanically. To determine all possible quantum domains, we need to determine the moduli $\mathcal{M}$ of the operator space $O = \sp{A_0 B_0, A_0 B_1, A_1 B_0, A_1 B_1}$.  This system has a non-trivial center as the operator $\sigma_y^a\sigma_y^b$ commutes with $O$.  By applying the approach through Eqs.~(\ref{eq:minor}--\ref{eq:elimination}), we obtain 
\begin{equation}\label{eq:bellf}
\begin{split}
\faf =  \Ave{A_0 B_0}^2+\Ave{A_0  B_1}^2+\Ave{A_1 B_0}^2+\Ave{A_1 B_1}^2 \\
    -\left(\Ave{A_0 B_0}\Ave{A_1 B_1}-\Ave{A_0 B_1}\Ave{A_1 B_0}\right)^2 -1,
\end{split}
\end{equation}
which provides a nonlinear bound $\faf \leq 0$ for the quantum expectations. It is important to note that Eq.~(\ref{eq:bellf}) contains spurious branches. To remove these unphysical branches and establish a direct connection with the Tsirelson bound~(\ref{eq:tsirelson}), it is enlightening to rewrite Eq.~(\ref{eq:bellf}) in terms of the following non-algebraic form
\begin{equation}\label{eq:bellb}
\begin{split}
&\sqrt{\frac{(\Ave{A_0 B_0}-\Ave{A_1 B_1})^2+(\Ave{A_0 B_1}+\Ave{A_1 B_0})^2}{2}} + \\
& \sqrt{\frac{(\Ave{A_0 B_0}+\Ave{A_1 B_1})^2+(\Ave{A_0 B_1}-\Ave{A_1 B_0})^2}{2}} \leq \sqrt{2},
\end{split}
\end{equation}
where the left side is expressed as a sum of two quadratic means (QM). The Tsirelson bound~(\ref{eq:tsirelson})  follows directly as $\frac{1}{2}| \Ave{A_0 B_0}+\Ave{A_0 B_1}+\Ave{A_1 B_0}-\Ave{A_1 B_1}| \leq \frac{1}{2}(|\Ave{A_0 B_0}-\Ave{A_1 B_1}|+|\Ave{A_0 B_1}+\Ave{A_1 B_0}|) \leq \sqrt{\frac{(\Ave{A_0 B_0}-\Ave{A_1 B_1})^2+(\Ave{A_0 B_1}+\Ave{A_1 B_0})^2}{2}} \leq \sqrt{2}$. The first inequality follows from the triangle inequality,  the second follows from the AM-QM inequality, and the third from Eq.~(\ref{eq:bellb}). Moreover, Eq.~(\ref{eq:bellb}) provides the best bounds, in the sense that any set of $\Ave{A_iB_j}$ values that satisfy it is allowed quantum mechanically, and any $\Ave{A_iB_j}$ values that violate it is disallowed. 

It is noteworthy that the boundary $ \partial{M}$ is mostly saturated by entangled states. Inquisitively, one might wonder about the expectation bounds for separable states. Using Eq.~(\ref{eq:sxy1}), one can obtain $\Ave{A_0}^2+\Ave{A_1}^2 \le 1$ and $\Ave{B_0}^2+\Ave{B_1}^2 \leq 1$, which implies that
\begin{equation}\label{eq:ss}
   \Ave{A_0 B_0}_\mathrm{ss}^2+\Ave{A_0 B_1}_\mathrm{ss}^2 + \Ave{A_1 B_0}_\mathrm{ss}^2 + \Ave{A_1 B_1}_\mathrm{ss}^2 \le 1.
\end{equation}
Here, $\Ave{A_iB_j}_\mathrm{ss} \coloneqq \Ave{A_i}\Ave{B_j}$ denotes the expectation values for separable states. In comparison to Eq.~(\ref{eq:bellb}), Eq.~(\ref{eq:ss}) provides a much weaker bound. Specifically, we can infer the Bell inequality~(\ref{eq:bell}) as $\frac{1}{4}|\Ave{A_0 B_0}+\Ave{A_0 B_1}+\Ave{A_1 B_0}-\Ave{A_1 B_1}| \leq \sqrt {\frac{\Ave{A_0 B_0}_\mathrm{ss}^2+\Ave{A_0 B_1}_\mathrm{ss}^2 + \Ave{A_1 B_0}_\mathrm{ss}^2 + \Ave{A_1 B_1}_\mathrm{ss}^2}{4}} \leq 1/2$. The first inequality follows from the AM-QM inequality, while the second arises from Eq.~(\ref{eq:ss}). In other words, the Bell bound establishes an upper limit for the expectation of separable states, while the Tsirelson bound sets an upper bound for the expectation of entangled states.

\begin{figure}
 \includegraphics[width=1\linewidth]{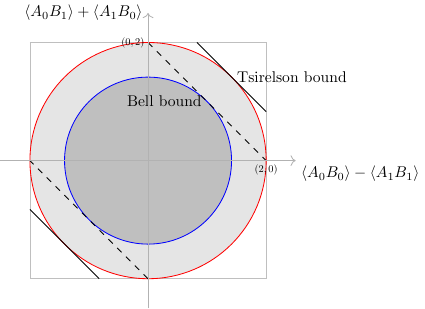}
 \caption{The moduli space $\mathcal{M}$ for operators $\Ave{A_0B_0}$, $\Ave{A_0B_1}$, $\Ave{A_1B_0}$ and $\Ave{A_1B_1}$, intersected with the hyperplanes $\Ave{A_0B_0} +\Ave{A_1B_1} = 0$ and $\Ave{A_0B_1}-\Ave{A_1B_0} = 0$, represented by the light grey disk (Eq.~(\ref{eq:bellb})). The dark grey disk (Eq.~(\ref{eq:ss})) represents the allowed domain for separable states. The dashed line corresponds to Bell bound, while the solid line corresponds to Tsirelson bound.}
 \label{fig:bell}
\end{figure}

Figure~\ref{fig:bell} illustrates a two-dimensional intersection of the moduli space $\mathcal{M}$ based on Eqs.~(\ref{eq:bellb})--(\ref{eq:ss}) and compares it with Eqs.~(\ref{eq:bell})--(\ref{eq:tsirelson}). We find that the Bell and Tsirelson bounds provide tangential lines of the moduli space for separable and entangled states, as expected. While our approach does not involve wavefunctions, the analysis above demonstrates its suitability for examining quantum nonlocality and entanglement. We leave the possibility of generalizing our method to other systems for future investigation. We argue that a theory that predicts equivalent expectation moduli for all physical observables is indistinguishable from quantum mechanics, as discussed in Subsec.~\ref{sec:cat}.

\subsection{Density functional theory}\label{sec:DFT}

\begin{figure*}
 \includegraphics[width=1\linewidth]{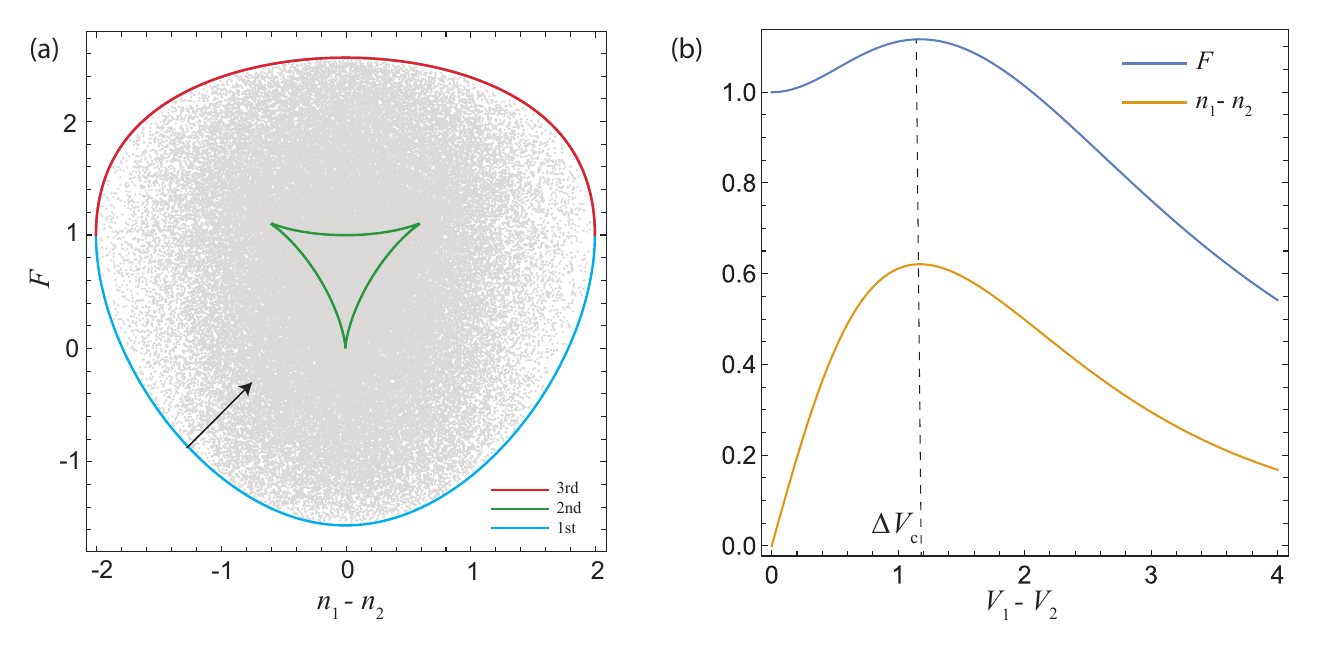}
 \caption{(a) The density functional $F[n]$ in unit of $t$ for the two-boson system for a fixed $U/t = 1$ and $U'/t = 0$. Curves are determined by Eq.~(\ref{eq:functional}), where colors correspond to three different eigenstates. Grey scatter points sample the moduli space $\mathcal{M}$, which is bounded by $F[n]$. The black arrow represents the normal vector $(2, V_1-V_2)$. (b) $F$ and $n_1-n_2$ as a function of $V_1-V_2$ for the first excited state.}
 \label{fig:q2n2}
\end{figure*}

In this subsection, we apply our approach to DFT for systems of particles subjected to an external potential $V$. The corresponding Hamiltonian is given by 
\begin{equation}\label{eq:DFT}
 H[V] =  H_0 + \int V(\mathbf{r}) \hat n(\mathbf{r}), 
\end{equation}
where $H_0$ represents the Hamiltonian containing the kinetic and pair interaction terms, and $\hat n(\mathbf{r})$ and $V(\mathbf{r})$ are the density operator and external potential at position $\mathbf{r}$, respectively. Similar to the Luttinger–Ward functional discussed in Subsec.~\ref{sec:FT}, Hamiltionian~(\ref{eq:DFT}) is parameterized by an uncountable set of parameters, $\lambda_0 = 1$, and $\lambda(\mathbf{r}) = V(\mathbf{r})$, over the operator space $O = \sp{H_0, \hat n(\mathbf{r})}$. Note that the identity operator has been included as the particle number operator $\hat N = \int \hat n(\mathbf{r})d\mathbf{r}$.

The moduli $\mathcal{M}$ of the expectation values $F \coloneqq \langle H_0 \rangle$ and $n(\mathbf{r}) \coloneqq \langle \hat n(\mathbf{r}) \rangle$ is an infinite-dimensional space. The corresponding singular moduli $\mathcal{M}_s$ are associated with the eigenstates of Eq.~(\ref{eq:DFT}), satisfying
\begin{equation}\label{eq:fDFT}
    \fpr(F, n(\mathbf{r})) = 0.
\end{equation}
Therefore, $F[n]$ can be viewed as an implicit functional of $n(\mathbf{r})$ and depends only on $H_0$. By substituting Eq.~(\ref{eq:E}), we obtain the energy functional
\begin{equation}\label{eq:HK}
E[n] = F[n] + \int V(\mathbf{r}) n(\mathbf{r}),
\end{equation}
which is bounded and uniquely saturated by the moduli boundary $\partial \mathcal{M}$ that corresponds to the ground state. Thus, our analysis  provides a constructive proof of Hohenberg and Kohn's (HK's) theorems \cite{hohenberg1964inhomogeneous}, where $F[n]$ is precisely the HK functional, and Equation~(\ref{eq:fDFT}) provides an explicit construction of the HK functional. Note that one may also rewrite Eq.~(\ref{eq:HK}) as 
\begin{equation}
    F[n] = E[V[n]] - \int V[n](\mathbf{r}) n(\mathbf{r}) d \mathbf{r},
\end{equation}
which provides the Legendre transform~(\ref{eq:Legendre}) of the dual functional $E[V]$, where $V[n]$ is determined from the Eq.~(\ref{eq:dfprd}), as 
\begin{equation}
    n(\mathbf{r}) = \frac{\delta E[V]}{\delta V(\mathbf{r})}.
\end{equation}

Below we demonstrate how our approach can be used to compute the exact density functional $F[n]$. For simplicity, we consider a system of $n$ identical particles with a single-particle Hilbert space $\mathcal{H}^1$ in a discrete space with a total of $q$ states. The system's Hilbert space is denoted as $\mathcal{H}^\pm = \mathrm{Sym}^\pm \otimes^n \mathcal{H}^1$, with dimensions $N = \binom{q+n-1}{n}$ and $\binom{q}{n}$ for bosons and fermions, respectively. Our Hamiltonian is given by
\begin{equation}\label{eq:DFT1}
 H =  H_0 + \sum_{i} V_i  \hat n_i, 
\end{equation}
where
\begin{equation}
 H_0  = \sum_{ij} \left(  -t_{ij} a_i^\dag a_j + U_{ij}  \hat n_i  \hat n_j \right).
\end{equation}
Here $a_i^\dag$ and $a_i$ are creation and annihilation operators, and $ \hat n_i \coloneqq a_i^\dag a_i $. Since $\sum n_i = n I$, the energy term $E$ is absorbed into the parameter set by $V_i \rightarrow V_i - E/n$. The corresponding parameter set $\lambda = (1, V_1, \ldots, V_q )$ and operator space  $O = \sp{H_0, n_1, \ldots, n_q }$ with $M = q$. The corresponding Jacobian is given by
\begin{equation}\label{eq:JDFT}
 \mathcal{J}_{DFT}(\psi, \psi^\dag)  \coloneqq 
 \begin{bmatrix}
 \psi^{\dag} H_0   , & \psi^t  H_0 ^t  \\
\psi^{\dag} \hat n_1   , & \psi^t  \hat  n_1 ^t  \\
\ldots, & \ldots  \\
\psi^{\dag} \hat n_q   , & \psi^t  \hat  n_q ^t
\end{bmatrix}.
\end{equation}
By using ideal computation of Eq.(\ref{eq:elimination}), we obtain the explicit form of Eq.(\ref{eq:fDFT}) as a homogeneous polynomial,
\begin{equation}\label{eq:functional}
\fpr(F,n_1 \ldots, n_q) = \sum_{d_0 + \ldots + d_q = d} c_{d_0,\ldots, d_q} n_1^{d_1} \ldots n_{q}^{d_q}  F^{d_0},
\end{equation}
where $d$ is the degree of the polynomial, and $c_{d_0,\ldots, d_q}$ are real constants depending only on $ H_0$. To demonstrate a concrete example of this construction, we set $t_{ij} = t$ for off-diagonal elements, $U_{ij} = U$ and $U'$ for diagonal and off-diagonal elements, respectively, and analytically solve Eq.~(\ref{eq:elimination}) for small $(n,q)$-systems for fixed parameters $t$, $U$, and $U'$.

We present a simple example of our approach by considering a system of $n=2$ bosons in a discrete space with two states ($q=2$), resulting in $N=3$ eigenstates. In this case, Eq.~(\ref{eq:functional}) has a degree of $d=6$ (see SM Eq.S1). The density functional $F[n]$ is a one-dimensional function since the total number of particles is fixed ($n_1 + n_2 = 2$). Figure~\ref{fig:q2n2}a illustrates the behavior of $F[n]$ by plotting it as a function of the difference $\Delta n = n_2 - n_1$. The plot shows three disconnected curves, corresponding to the three eigenstates of the system, where the normal vector corresponds to $(2, V_1-V_2)$. Note that the ground state and the highest excited state are smoothly connected, as expected.

To test our theory, we randomly generate the wavefunction $\psi$ over $100,000$ instances and numerically calculate the corresponding expectation values. This approach enables us to sample the moduli space $\mathcal{M}$ numerically, as depicted by the gray scatter points in Fig. \ref{fig:q2n2}. Our findings reveal that the moduli space $\mathcal{M}$ is precisely bounded by the ground state moduli (blue curve), which is in agreement with our theoretical predictions. Furthermore, we use exact diagonalization of Hamiltonian~(\ref{eq:DFT1}) to compute eigenfunctions directly and find that the result is consistent with the curves generated by Eq.~(\ref{eq:functional}).

To better understand the ground state moduli, we find that the lowest functional 
\begin{equation}
  F_0 = U - \frac{1}{2} \left(\Delta U + \sqrt{\Delta U ^2+(4t)^2}\right),  
\end{equation}
with $\Delta U = U - U'$ occurs at $n_1 = n_2$, corresponding to the normal vector $(1,0)$, i.e., $V_1 = V_2$. Increasing $V_1$ or $V_2$ drives the ground state to a higher value of $F[n]$. Expanding $F[n]$ around $F_0$ in terms of the power series of $n_1n_2$, we obtain $F[n] = F_0 + F_1 n_1 n_2 + O((n_1 n_2)^2)$, where
\begin{equation}
  F_1 =    \frac{4 t^2 \sqrt{\Delta{U}^2+16 t^2}}{16 t^2-2 \Delta{U} \left(\sqrt{\Delta{U}^2+16 t^2}-\Delta{U}\right)}.
\end{equation}
For weak coupling $\Delta U/t \ll 1$, we have $F_1 = - t^2 - \Delta U/2 + O(\Delta U ^2)$. This can be interpreted as the kinetic and the mean-field interaction terms. However, the exact interaction expectation $\langle \hat U \rangle = U_1-\Delta U \langle \hat n_1 \hat n_2 \rangle$ has a different prefactor, reflecting the underlying correlation $\langle \hat n_1 \hat n_2 \rangle \neq \langle \hat n_1 \rangle \langle \hat n_2 \rangle$.

The excited state exhibits a more complex geometry than the ground state, as depicted by the green curve in Fig.~\ref{fig:q2n2}a. The central cusp at $n_1=n_2$ corresponds to the limiting cases $V_1 \rightarrow \infty$ or $V_2 \rightarrow \infty$. To gain further insight into the cusps at two sides, we plot $F$ and $n_1-n_2$ as functions of the normal vector $V_1 - V_2$ in Fig.~\ref{fig:q2n2}b, which reveals that both quantities peak at $V_1-V_2=\Delta V_c \approx 1.2$. By expanding $\delta F = \frac{1}{2} F''(\Delta V_c) \delta V^2 + \frac{1}{6} F'''(\Delta V_c) \delta V^3 + O(\delta V^4)$ and $\delta \Delta n= \frac{1}{2} \Delta n''(\Delta V_c) \delta V^2 + \frac{1}{6} \Delta n'''(\Delta V_c) \delta V^3 + O(\delta V^4)$ around $\Delta V_c$, we obtain the standard cusp equation $y^2 = x^3$, where $x$ and $y$ are linear combinations of $\delta F$ and $\delta \Delta n$.

We now extend our analysis to a system of $n=3$ particles. By following a similar approach as before, we obtain Eq.~(\ref{eq:functional}) with a $d=12$ degree polynomial (see SM Eq.S2). Figure~\ref{fig:q2n3} illustrates the behavior of the functional $F[n]$ by plotting it as a function of $n_2 - n_1$. In this system, there are four eigenstates, each corresponding to the colored segments of the $F(n)$ curve. Notably, the ground state and the highest excited state are connected, and the two middle excited states are also connected due to their projective nature. Moreover, these two excited states exhibit more singularities than the $n=2$ case, with six cusps and three crunodes. The top four cusps occur when both $F$ and $n_1 -n_2$ reach their maximum values simultaneously, similar to the left and right cusps observed in the $q=2$ case. The bottom two cusps occur at the limiting case when $V_1$ or $V_2$ diverges. This is also the case when these two excited states degenerate and the two corresponding curves join together. The middle crunode appears when the excited state has $n_1 = n_2$ while $V_1 \neq V_2$. The exchange of $V_1$ and $V_2$ leads to the same functional $F[n]$, due to the symmetry. The other two crunodes occur when different excited states share the same values of $F$ and $n_1 -n_2$.

\begin{figure}
\centering
 \includegraphics[width=1\linewidth]{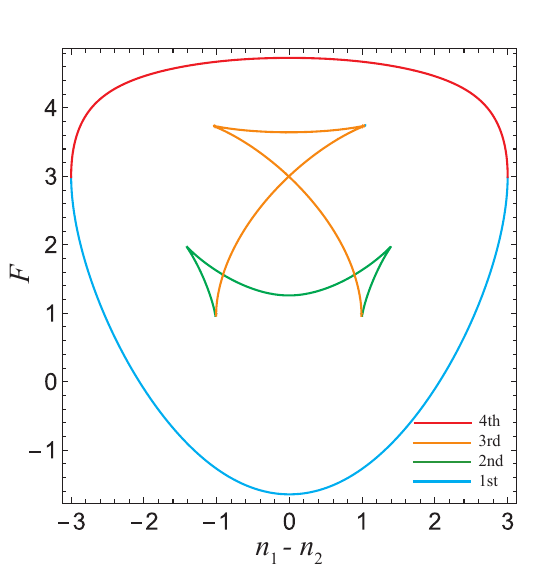}
 \caption{The density functional $F[n]$ for the three-boson system for a fixed $U/t = 1$ and $U'/t = 0$. 
 }
 \label{fig:q2n3}
\end{figure}

Finally, we apply our analysis to a system of $q=3$ states, where the density functional $F[n]$ is a two-dimensional surface depicted in Fig.~\ref{fig:q3n2}, since there are two free variables. The domain enclosed by the yellow surface is quantum mechanically allowed, where the ground states correspond to the bottom surface. Notably, as $V_i \to \infty$ and $n_i=0$ for state \emph{i}, the high-dimensional density functional reduces to the low-dimensional case depicted in Fig.~\ref{fig:q2n2}a, which imposes a marginal constraint on $F[n]$. For a general framework of this projection, readers can refer to Subsec.~\ref{sec:cat}.  Our observation suggests a potential possibility to reconstruct $F[n]$ approximately from the low-dimensional margins as an optimal transport problem \cite{villani2021topics}, given an appropriate cost function. We leave this possibility for future investigation.

\begin{figure}
\centering
 \includegraphics[width=1\linewidth]{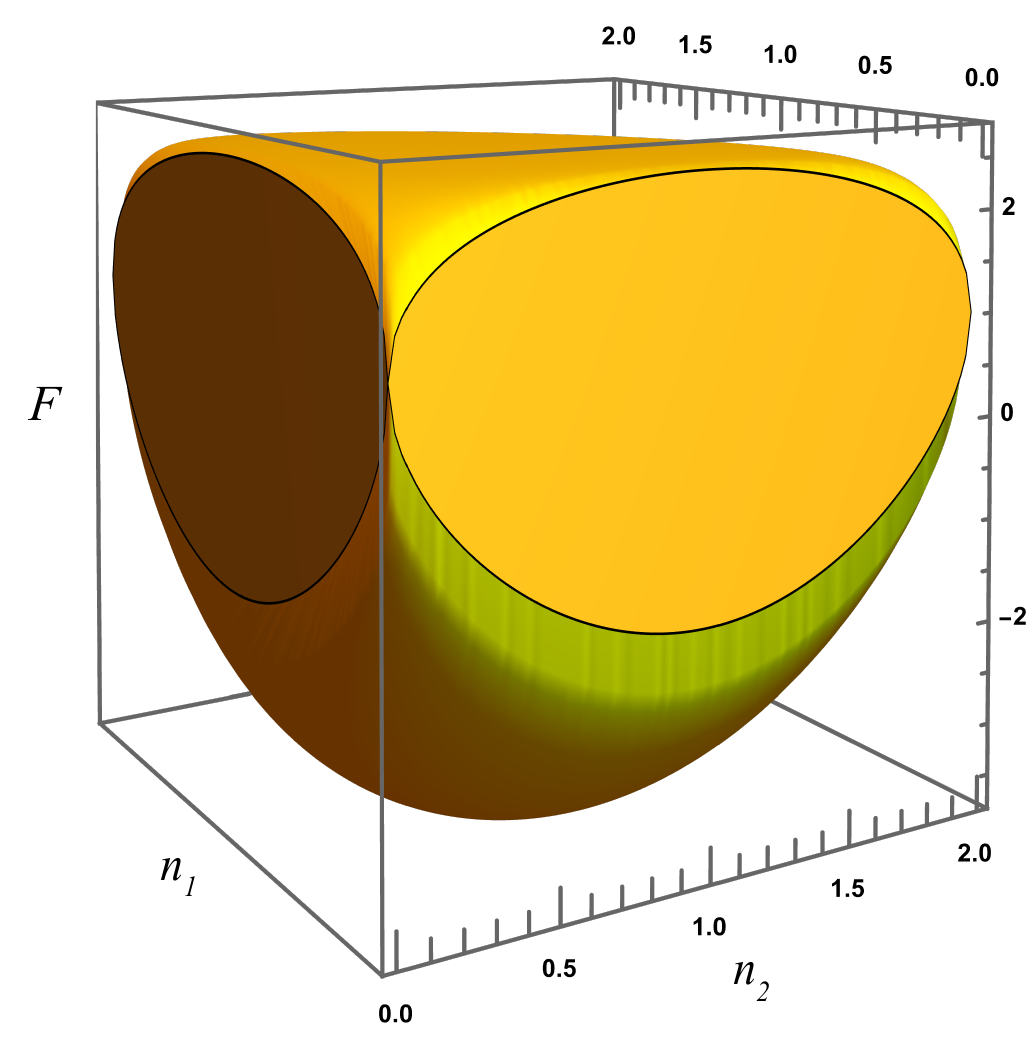}
 \caption{The moduli space $\mathcal{M}(F,n_1,n_2)$ for the $q=3$ state system for a fixed $U/t = 1$ and $U'/t = 0$. The marginal case $n_1 = 0$ or $n_2 = 0$ is equivalent to the moduli space shown in Fig.~\ref{fig:q2n2}a.}
  \label{fig:q3n2}
\end{figure}

The above analysis provides illustrative examples of how our approach can be utilized to derive exact density functionals, potentially offering insight into the intricate geometry of density functionals, which is reflected in the explicit formula even for the two-site case. Although lattice-type models may not perfectly represent real-space DFT problems, they provide a valuable opportunity to investigate the deviation of existing DFT approximations from the exact solution. Our approach allows us to explore the contributions of non-local terms, often not fully understood in existing approximations, thereby guiding the design of improved approximations for problems in continuous space.

On the other hand, the complexity of the geometry of density functions increases rapidly with system size, as finding the exact density functional is generally an NP-hard problem. While our current na\"ive algorithm is limited to very small systems, there are no fundamental restrictions on our construction. Similar to exact diagonalization, it is possible to develop numerical algorithms capable of handling reasonably small-sized systems. We will explore such opportunities in future research.

The Pl\"ucker formula~(\ref{eq:plucker}) implies that the degree of $\fpr$ decreases with the number of singularities on its dual hypersurface $\mathcal{M}^*$. Thus, if $\mathcal{M}^*$ possesses a large set of symmetries, i.e., a significant number of singularities, then the complexity of $\fpr$ can be greatly reduced. However, the symmetry of DFT is limited by the chosen $H_0$ operator, resulting in a complex geometric structure. To overcome this issue, one possible solution is to expand the Hamilton family with extra symmetries. A potential approach is to replace the particle density with $2$-particle reduced density matrices ($2$-RDMs), as most physically relevant problems involve only two-body interactions. This approach eliminates the dependence on a specific $H_0$ operator, allowing the geometry of $F[n]$ to be transferred to the geometry of RDMs, which hopefully has a simpler structure \cite{mazziotti2007reduced,coleman2000reduced,mayer1955electron}. We will discuss this further in the next subsection.

\subsection{ $N$-representability conditions} \label{sec:NRP}

In this subsection, we briefly discuss the moduli space of the reduced density matrices, also known as the $N$-representability conditions, which we explain in detail below. We begin with the general many-body Hamiltonian that involves $m$-body interactions, defined as
\begin{equation}\label{eq:HRDM}
    H(h^{(m)}) =  \sum_{i_1\ldots i_m; j_1 \ldots j_m} h^{(m)}_{i_1 \ldots i_m; j_1\ldots j_m} a_{i_1}^\dag\ldots a_{i_m}^\dag  a_{j_1}\ldots a_{j_m}, 
\end{equation}
where $a$ and $a^\dag$ are the annihilation and creation operators, respectively, and the $m$-particle reduced Hamiltonian $h^{(m)}$ is self-adjoint. Practically, $m=2$ is sufficient since most systems involve only pairwise interactions. The problem of finding an $N$-particle ground energy of Eq.~(\ref{eq:HRDM}) can be transformed into a low-dimensional optimization problem of the following energy functional \cite{mazziotti2007reduced,coleman2000reduced,mayer1955electron},
\begin{equation}\label{eq:Erho}
    E[\rho^{(m)}] \coloneqq \ave{H} = \mathrm{tr}( h^{(m)}\rho^{(m)}),
\end{equation}
where $\rho^{(m)}$ is $m$-particle reduced density matrix ($m$-RDM), defined as
\begin{equation}\label{eq:RDM}
    \rho^{(m)}_{i_1, \ldots, i_m; j_1, \ldots, j_m} \coloneqq \ave{a_{i_1}^\dag\ldots a_{i_m}^\dag  a_{j_1}\ldots a_{j_m} }.
\end{equation}
In Eq.~(\ref{eq:Erho}), the trace is taken over the reduced Hilbert space with $m$-particles, which simplifies significantly the original many-body problem~(\ref{eq:HRDM}).  However, the space of all $m$-RDMs often has a highly non-trivial geometry. Direct optimization of Eq.~(\ref{eq:Erho}) without considering this geometry typically results in finding the wrong ground state. Therefore, determining the conditions necessary for all possible $m$-RDMs, namely, the $N$-representability conditions, becomes a challenging problem in computational quantum chemistry \cite{mayer1955electron, coleman1963structure,harriman1978geometry,erdahl1978representability,erdahl1979two,percus1978role,kryachko2014density}.

 Our general framework suggests that the space of all $m$-RDMs is the moduli space $\mathcal{M}^{(m)}$ of expectation values as defined in Eq.~(\ref{eq:RDM}), and the $N$-representability conditions can be translated into finding corresponding $\fpr$ and $\partial \mathcal{M}^{(m)}$. We can choose a set of self-adjoint bases such that the coefficient of $h^{(m)}$ is real. Below we present our results mainly in a coordinate-free form, so the choice of the basis will not affect our findings.

Using equation~(\ref{eq:fprd}), the dual singular moduli can be written explicitly as
\begin{equation}\label{eq:dfRDM}
    \fpr^*(h^{(m)}) = \det \left(H(h^{(m)})\right),
\end{equation}
where the reduced Hamiltonian $h^{(m)}$ is the dual variable, and the determinant is taken over the many-body particle Hilbert space.  Although finding the dual to Eq.~(\ref{eq:dfRDM}) may seem as difficult as solving the original many-body problem~(\ref{eq:HRDM}), at least for the case of $m=1$, the singular moduli $\mathcal{M}^{(1)}_e$ can be determined by a simple relation,
\begin{equation}\label{eq:fRMD}
    \fpr(\rho^{(1)}) = \det(\rho^{(1)}) = 0,
\end{equation}
where the determinant is taken over the one-particle reduced Hilbert space. This suggests that $\mathcal{M}^{(1)}$ is the space of all non-negative definite matrices. In other words, any non-negative definite matrix can be a single-particle RDM. 

To prove Eq.~(\ref{eq:fRMD}), we only need to show the dual transformation of Eq.~(\ref{eq:fRMD}) vanishes Eq.~(\ref{eq:dfRDM}). Using Eq.~(\ref{eq:dfpr}), we find that $h^{(1)}$ is proportional to the adjoint matrix of $\rho^{(1)}$, i.e., $h^{(1)} \sim \mathrm{adj}(\rho^{(1)})$. Equation~(\ref{eq:fRMD}) implies that $\rho^{(1)}$ has at least corank one. If its corank is greater than one, then $\mathrm{adj}(\rho^{(1)}) = 0$, which vanishes Eq.~(\ref{eq:dfRDM}) trivially. If it has corank one, then its adjoint matrix  $\mathrm{adj}(\rho^{(1)})$ has rank one. This implies $h^{(1)} = | v \rangle\langle v |$, for some complex vector $|v\rangle$. Substituting into Eq.~(\ref{eq:HRDM}), we find $H(h^{(1)}) = \tilde a^\dag \tilde a$, where $\tilde a \coloneqq \sum_i v_i a_i$, which vanishes the determinant in Eq.~(\ref{eq:dfRDM}) for systems with more than one state. This completes our proof of Eq.~(\ref{eq:fRMD}).

While an explicit expression for $\fpr^{(m)}$ is currently unknown for $m>1$, it may be possible to derive it computationally for small systems using our approach. However, as the $N$-representability conditions are known to be NP-hard \cite{deza1997geometry}, the degree of $\fpr$ will grow rapidly with the system size, making it increasingly difficult to obtain explicit expressions for larger systems. Nevertheless, it is possible that observations of small systems may provide valuable insights into the structure of $\fpr^{(m)}$ and guide the search for a general expression, perhaps analogous to the simple relation for $m=1$ in Eq.~(\ref{eq:fRMD}). Alternatively, one could explore connections between our approach and existing techniques such as tensor decompositions of a set of model Hamiltonians \cite{mazziotti2012structure}, which provides sequential linear approximations to $\mathcal{M}^{(2)}$. We leave the pursuit of these possibilities to future research.

\section{Conclusion}
Understanding the nature of quantum theory has been a fundamental and critical question since its inception. Various quantum formulations have been proposed to answer this question, each providing different perspectives on this subject. In this paper, we offer a new perspective on time-independent quantum theory by shifting the focus from the quantum states in Hilbert space to the quantum geometry of expectation values. We develop a general framework for quantum geometry over an arbitrary set of physical observables and establish its connection to eigenstates. Moreover, these geometries can be viewed as a ``quantization" of their classical counterparts.

One of the key findings of our theory is that the boundary of expectation value moduli provides a natural quantum bound to physical observables, with the Heisenberg uncertainty relation being one of its special cases. This result opens up a vast generalization of the uncertainty principle, providing a new and exciting direction for exploring the foundations of quantum theory.

Our approach leaves many challenges and opportunities for future research. For example, extending our framework to a time-dependent theory would complete our alternative formulation of quantum theory. There are several possible avenues to pursue this extension. The simplest way is to connect to the path-integral method proposed in Subsec.~\ref{sec:FT}, although this may come at the cost of losing the elegance of our geometric approach. Another option is to model the time evolution of expectation value moduli as a time-dependent projection, as discussed in Subsec.~\ref{sec:cat}. Alternatively, we could explore the possibility of generalizing our approach using the time-dependent variation principle~\cite{kramer1980geometry}. Moreover, an important question is whether Hilbert space can be reconstructed from the proposed expectation value category. This question is closely related to the recently developed geometrical quantum formulation \cite{kibble1979geometrization, ashtekar1995geometry}.

In addition to the challenges posed, our framework presents numerous opportunities for application in diverse fields. To showcase its versatility, we present several applications. Notably, we derive a new nonlinear quantum bound that violates the Bell inequality, which is stronger than the existing Tsirelson bound. Additionally, we demonstrate that our framework encompasses density functional theory by providing an explicit construction of the HK functional. We derive analytical forms of exact density functionals in small systems and analyze their associated geometries. These results offer insights into the exact density functional, potentially leading to a better design for future density functional approximations. Furthermore, we discuss how our theory connects to the long-standing challenge of determining the geometry of reduced density matrices, known as the $N$-representability conditions in computational quantum chemistry.

Moreover, we want to briefly mention another application of our theory to understanding quantum criticality, which is not discussed in this paper. We have discovered that the quantum critical point corresponds to the zero curvature of the moduli boundary, which provides an intriguing connection between quantum criticality and moduli geometry. Readers can find more details in Reference~\cite{song2023zero}. Overall, our approach offers a novel perspective on quantum theory with potential implications in diverse fields,  such as quantum entanglement, strongly correlated systems, and quantum computational chemistry. We hope that our work will stimulate further research in these areas, leading to new insights into the nature of quantum theory.

% \begin{acknowledgments}
% C.S. was supported partially by the National Science Foundation under Grants 2150830 and IBSS-1620294, the Institute of Education Sciences under Grant R324A180203, and the National Institutes of Health under Grant R01DC018542. N.F.J. was funded by AFOSR grants FA9550-20-1-0382 and FA9550-20-1-0383.

% \end{acknowledgments}

%\bibliography{ref}% Produces the bibliography via BibTeX.

%

\end{document}